\documentclass[aps,prd,preprint,groupedaddress]{revtex4-2}
\usepackage{amsmath}
\usepackage{tikz}
\usepackage[colorinlistoftodos]{todonotes}
\definecolor{blau}{RGB}{0,15,117}
\definecolor{grana}{RGB}{138,18,20}
\usetikzlibrary{decorations.pathmorphing}
\usetikzlibrary{decorations.markings}
\usetikzlibrary{positioning, shapes, snakes, arrows}
\tikzset{
gluon/.style={decorate,
 decoration={coil,amplitude=2pt, segment length=2pt,  pre length=.1cm, post length=.1cm}}
}

\begin{document}


\title{Medium evolution of a static quark-antiquark pair in the large $N_{c}$
limit}


\author{Miguel \'{A}ngel \surname{Escobedo}}
\email[]{miguelangel.escobedo@usc.es}
\affiliation{Instituto Galego de Física de Altas Enerx\'{i}as (IGFAE), Universidade
de Santiago de Compostela, E-15782, Galicia, Spain.}


\date{\today}

\begin{abstract}
We study the transitions between the different color states of a static
quark-antiquark pair, singlet and octet, in a thermal medium. This
is done non-perturbatively exploiting the infinite mass limit of QCD. This study is interesting because it can be used
for future developments within the framework of Effective Field Theories
(EFTs) and because it can be combined with other techniques, like lattice
QCD or AdS/CFT, to gain non-perturbative information about the evolution
of quarkonium in a medium. We also study the obtained expressions
in the large $N_{c}$ limit. This allows us to learn lessons that are
useful to simplify phenomenological models of quarkonium in a plasma.
\end{abstract}


\maketitle

\section{Introduction}
Heavy quarkonium suppression was proposed in \cite{Matsui:1986dk}
as a probe of the formation of a quark-gluon plasma. Since then, it
has been intensively studied, both theoretically and experimentally.
Quarkonium suppression has been observed at the LHC and RHIC (see \cite{Ma:2018tmg}
for a recent review). However, the original goal of using quarkonium
to obtain relevant information about the medium is not yet completely
fulfilled.

On the theoretical side, three main mechanisms have been identified
as relevant to understand the suppression pattern of quarkonium: screening,
thermally induced decay into octets and regeneration. Screening is
the original mechanism studied in \cite{Matsui:1986dk}. In a deconfined
medium, static chromoelectric fields are screened at distances of
the order of the Debye radius or bigger. This implies that, if the
typical size of a particular bound state is much bigger than the Debye
radius, then the bound state cannot exist in such a medium. Another
mechanism that can produce the dissociation of quarkonium is the existence
of a thermally induced decay width. This process started to gain attention
after the computation in \cite{Laine:2006ns}. The mechanism responsible for this imaginary part is the medium induced transition from color singlet to color octet \cite{Brambilla:2008cx}. Since then, many works have studied this mechanism and have identified it as important in a wide range of temperature regimes \cite{Beraudo:2007ky,Brambilla:2008cx,Escobedo:2008sy,Brambilla:2010vq}.
Finally, recombination \cite{BraunMunzinger:2000px,Thews:2000rj}
is a process that can counterbalance suppression and in which two initially
uncorrelated heavy quarks recombine inside the medium to form a new
bound state. This effect might be relevant in charmonium
since it is abundantly produced in heavy-ion collisions at the LHC. Indeed,
some models need the inclusion of this effect in order to reproduce
experimental data \cite{Zhao:2011cv,Du:2015wha}.

A theoretical framework in which all these effects can be included
in a consistent way is that of open quantum systems \cite{Borghini:2011ms,Akamatsu:2012vt,Akamatsu:2014qsa,Akamatsu:2018xim,Brambilla:2016wgg,Brambilla:2017zei,Kajimoto:2017rel,Katz:2015qja,Yao:2018nmy,Yao:2018sgn,Blaizot:2017ypk,Blaizot:2018oev,Blaizot:2015hya}.
The heavy-quark pair is treated as an open system interacting with
an environment formed by light quarks and gluons. The reduced density
matrix of heavy quarks evolves following a master equation, which
in the Markovian limit takes the form of a Gorini-Kossakowski-Sudarshan-Lindblad (GKSL) equation \cite{Gorini:1975nb,Lindblad:1975ef}.
Examples of phenomenological applications of this approach which compare with
experimental data can be found in \cite{Brambilla:2016wgg,Brambilla:2017zei,Brambilla:2020qwo}.

At the moment, most of the derivations of the master equation have
been done using perturbative arguments \cite{Akamatsu:2012vt,Akamatsu:2014qsa,Blaizot:2015hya,Blaizot:2017ypk,Blaizot:2018oev,Yao:2018nmy}
or assuming that all the thermal scales are much smaller than $\frac{1}{r}$
\cite{Brambilla:2016wgg,Brambilla:2017zei} ($r$ being the typical
radius). In this article, we would like to contribute toward a non-perturbative
understanding by studying the problem in the static limit. This limit
has already been used in the past to clarify some of the features
of the framework of open quantum systems \cite{Brambilla:2017zei,Blaizot:2018oev}. We would also
like to note that, in the $T=0$ case, using the EFT potential
non-relativistic QCD (pNRQCD) \cite{Pineda:1997bj,Brambilla:1999xf,Brambilla:2004jw}
it is possible to use the potential computed in the static limit as
an ingredient for computations in which the heavy quark mass is keep
finite. In this paper, we will argue that this is also true at finite temperature, although we will not prove it rigorously. A rigorous proof would probably involve the development of a new version of pNRQCD valid for the study of the evolution of the density matrix in the $\frac{1}{r}\sim T$ regime and a precise connection between the pNRQCD singlet and octet fields and the interpolating fields that we use in this manuscript. However, we leave that for future developments within the EFT framework

Note that our derivation in the static limit is valid for any relation between the particles' typical energy in the medium and the separation between the heavy quarks. No relation between the time scales present in the problem is assumed. It is only in section \ref{sec:conn}, in order to connect with heavy quarkonium physics, that this kind of assumption is needed.

On top of this, we will also use the large $N_{c}$ limit in order
to gain insight into the non-perturbative expressions that we obtained. This will allow us to gain some useful phenomenological insights, mainly the following:
\begin{itemize}
\item Heavy quarks in an octet state evolve in time as uncorrelated particles.
\item The rate of decay of an octet into a singlet is suppressed by a power of $\frac{1}{N_{c}^{2}}$.
\item The survival probability of a singlet can be described with a modification
of the effective action. 
\end{itemize}

The outline of the article is as follows. In section \ref{sec:Conditional-probabilities-in},
we discuss the conditional probabilities to find a static quark-antiquark
pair in a given color state at time $t$ provided that we know the
color state at some other time $t_{0}$. In section \ref{sec:conn}, we argue that these results are relevant to the physics of heavy quarkonium. In section \ref{sec:largenc}, we discuss what can be deduced from the use of the Fierz identity and the large $N_{c}$ limit. An alternative graphical representation that is useful to study the classical limit is given in section \ref{sec:class}. Next, in section \ref{sec:Comparison-with-perturbative}, we discuss whether previous computations in the literature agree in the large $N_c$ limit with what we discussed in section \ref{sec:Results-in-the}. Finally, we give our conclusions in section \ref{sec:Conclusions}.
\section{Conditional probabilities in the static limit\label{sec:Conditional-probabilities-in}}

As a starting point, we consider non-relativistic QCD (NRQCD) \cite{Caswell:1985ui,Bodwin:1994jh} in the limit in which the heavy quarks have an infinite mass

\begin{equation}
\mathcal{L}=\mathcal{L}_{QCD}+\psi^{\dagger}iD_{0}\psi+\chi^{\dagger}iD_{0}\chi\,,\label{eq:Lagrangian}
\end{equation}

where $\mathcal{L}_{QCD}$ is the QCD Lagrangian including as degrees
of freedom just gluons and light quarks, $\psi$ is the field that
annihilates an infinitely heavy quark and $\chi$ is the field that
creates an infinitely heavy antiquark. With these fields, we try to
create a static quark-antiquark pair that transforms as a color singlet.
Taking guidance from the interpolating fields discussed in \cite{Brambilla:2004jw},
we use 
\begin{eqnarray}
&S({\bf R},{\bf r},t)=\nonumber\\
&\frac{1}{\sqrt{N_{c}}}\chi^{\dagger}\left({\bf R}-\frac{{\bf r}}{2},t\right)\phi\left({\bf R}-\frac{{\bf r}}{2},{\bf R}+\frac{{\bf r}}{2};t\right)\psi\left({\bf R}+\frac{{\bf r}}{2},t\right)\,,
\label{eq:Sinter}
\end{eqnarray}
where 

\begin{equation}
\phi({\bf y},{\bf x};t)=P\,\exp\left\{ -i\int_{0}^{1}\,ds({\bf y}-{\bf x})g{\bf A}({\bf x}-s({\bf x}-{\bf y}),t)\right\} \,.
\end{equation}

Note, however, that the field $S$ is just a combination of NRQCD operators and not the singlet field of pNRQCD. The latter can be related under certain conditions and approximations with the field defined in eq. (\ref{eq:Sinter}) by the matching procedure. The normalization is chosen such that if we have a state (in the Heisenberg picture)

\begin{equation}
|M\rangle=\int\,d^{3}R\,d^{3}r\Psi({\bf R},{\bf r})S^{\dagger}({\bf R},{\bf r})|\Omega\rangle\,,\label{eq:ws}
\end{equation}

where $|\Omega\rangle$ is a normalized state of the theory without heavy
quarks, then $\langle M|M\rangle=1$ such that $\int\,d^{3}R\,d^{3}r|\Psi({\bf R},{\bf r})|^{2}=1$
is always fulfilled. We can proceed similarly in the case of an octet \footnote{Again; this is not directly related to the octet field in pNRQCD.}
\begin{equation}
O^{A}({\bf R},{\bf r},t)=\sqrt{2}\chi^{\dagger}\left({\bf R}-\frac{{\bf r}}{2},t\right)\phi\left({\bf R}-\frac{{\bf r}}{2},{\bf R};t\right)T^{A}\phi\left({\bf R},{\bf R}+\frac{{\bf r}}{2};t\right)\psi\left({\bf R}+\frac{{\bf r}}{2},t\right)\,.
\end{equation}
Now the normalization is chosen such that if we have a state

\begin{equation}
|M'\rangle=\int\,d^{3}R\,d^{3}r\Psi^{'}({\bf R},{\bf r})O^{A\dagger}({\bf R},{\bf r})|\Omega^{A}\rangle\,,\label{eq:wo}
\end{equation}

where $|\Omega^{A}\rangle$ is a state of the theory without heavy
quarks chosen such that $|M'\rangle$ fulfills Gauss's law and
with the property $\langle\Omega^{A}|\Omega^{B}\rangle=\frac{\delta^{AB}}{N_{c}^{2}-1}$,
then $\langle M'|M'\rangle=1$ such that $\int\,d^{3}R\,d^{3}r|\Psi'({\bf R},{\bf r})|^{2}=1$
is always fulfilled.

The probability of finding the static
pair in a singlet configuration is 

\begin{equation}
P_{s}({\bf R},{\bf r},t)=Tr(S^{\dagger}({\bf R},{\bf r},t)S({\bf R},{\bf r},t)\rho)\,.
\end{equation}

Equivalently, for the octet we have

\begin{equation}
P_{o}({\bf R},{\bf r},t)=Tr(O^{A\dagger}({\bf R},{\bf r},t)O^{A}({\bf R},{\bf r},t)\rho)\,.\label{eq:Po}
\end{equation}

These equations are true for any density matrix $\rho$ representing a state with one static quark and one static antiquark. In particular, they are valid for a linear combination of pure state density matrices in which the states are of the form of eqs. (\ref{eq:ws}) and (\ref{eq:wo}). Using the Fierz identity, we can prove that
\begin{equation}
\begin{split}
& S^{\dagger}({\bf R},{\bf r},t)S({\bf R},{\bf r},t)+O^{A\dagger}({\bf R},{\bf r},t)O^{A}({\bf R},{\bf r},t)\\
&=\psi^{\dagger}\left({\bf R}+\frac{{\bf r}}{2},t\right)\psi\left({\bf R}+\frac{{\bf r}}{2},t\right)\chi\left({\bf R}-\frac{{\bf r}}{2},t\right)\chi^{\dagger}\left({\bf R}-\frac{{\bf r}}{2},t\right)\,.
\end{split}
\label{eq:on}
\end{equation}
The operator on the second line leaves unchanged states which have one
heavy quark and one heavy antiquark. Using this, we can prove that,
in a suitably normalized state, it is fulfilled that 

\begin{equation}
\int\,d^{3}r\,d^{3}R\left(P_{s}({\bf R},{\bf r},t)+P_{o}({\bf R},{\bf r},t)\right)=1\,.
\end{equation}

We can generalize the results slightly by introducing the reduced density matrix projected to the singlet or the octet state
\begin{subequations}
\begin{equation}
\rho_s(\mathbf{R},\mathbf{r},\mathbf{R}',\mathbf{r}',t)=Tr(S^{\dagger}({\bf R}',{\bf r}',t)S({\bf R},{\bf r},t)\rho)\,,
\end{equation}
\begin{equation}
\begin{split}
\rho_o(\mathbf{R},\mathbf{r},\mathbf{R}',\mathbf{r}',t)=\\
Tr(O^{B\dagger}({\bf R}',{\bf r}',t)W^{BA}(\mathbf{R}',\mathbf{R},t)O^A({\bf R},{\bf r},t)\rho)\,,
\end{split}
\end{equation}
\end{subequations}
respectively, where $W^{AB}$ is a Wilson line in the adjoint representation\\ $W^{AB}({\bf y},{\bf x};t)=P\,\exp\left\{- \int\,ds({\bf y}-{\bf x})gf^{CAB}{\bf A}^{C}({\bf x}-s({\bf x}-{\bf y}),t)\right\}$. 

Now let us look at eq. (\ref{eq:Lagrangian}); we can introduce the
field redefinitions

\begin{align}
\psi({\bf x},t)= & U(t,-\infty;{\bf x})\tilde{\psi}({\bf x},t)\,,\\
\chi({\bf x},t)= & U(t,-\infty;{\bf x})\tilde{\chi}({\bf x},t)\,,
\end{align}

where $U(t,t_{0};{\bf x})=T\,\exp\left\{ i\int_{t_{0}}^{t}\,dt'gA_{0}({\bf x},t')\right\} $.
Using this, we can rewrite 

\begin{equation}
\mathcal{L}=\mathcal{L}_{QCD}+\tilde{\psi}^{\dagger}i\partial_{0}\tilde{\psi}+\tilde{\chi}^{\dagger}i\partial_{0}\tilde{\chi}\,.
\end{equation}

The tilde fields do not interact with the rest of the particles
and have a trivial equation of motion. We can use this to conclude
that

\begin{equation}
\psi({\bf x},t)=U(t,t_{0};{\bf x})\psi({\bf x},t_{0})\,,
\end{equation}

\begin{equation}
\chi({\bf x},t)=U(t,t_{0};{\bf x})\chi({\bf x},t_{0})\,.
\end{equation}

Applying this formula to the singlet and octet fields, we get
\begin{equation}
\begin{split}
S({\bf R},{\bf r},t)&=\frac{1}{\sqrt{N_{c}}}\chi^{\dagger}\left({\bf R}-\frac{{\bf r}}{2},t_{0}\right)U^{\dagger}\left(t,t_{0};{\bf R}-\frac{{\bf r}}{2}\right)\times\\
&\times\phi\left({\bf R}-\frac{{\bf r}}{2},{\bf R}+\frac{{\bf r}}{2};t\right)U\left(t,t_{0};{\bf R}+\frac{{\bf r}}{2}\right)\psi\left({\bf R}+\frac{{\bf r}}{2},t_{0}\right)\,,
\end{split}
\end{equation}

and

\begin{equation}
\begin{split}
O^{A}({\bf R},{\bf r},t) & = \sqrt{2}\chi^{\dagger}\left({\bf R}-\frac{{\bf r}}{2},t_{0}\right)U^{\dagger}\left(t,t_{0};{\bf R}-\frac{{\bf r}}{2}\right)\times\\
&\times\phi\left({\bf R}-\frac{{\bf r}}{2},{\bf R};t\right)T^{A}\phi\left({\bf R},{\bf R}+\frac{{\bf r}}{2};t\right)U\left(t,t_{0};{\bf R}+\frac{{\bf r}}{2}\right)\psi\left({\bf R}+\frac{{\bf r}}{2},t_{0}\right)\,.
\end{split}
\end{equation}
A word of caution regarding the time ordering is needed. We are going
to use the Schwinger-Keldysh contour in the following derivation. In this framework, the correct time ordering
is implemented  according to the branch of the contour to which the field belongs. The order of the operators in the formulas illustrates how the color indices are multiplied. Now we can use the Fierz identity
to derive the following relation

\begin{equation}
\begin{split}
\delta_{aa'}\delta_{b'b} & =2\left(\phi\left({\bf R}-\frac{{\bf r}}{2},{\bf R};t\right)T^{A}\phi\left({\bf R},{\bf R}+\frac{{\bf r}}{2};t\right)\right)_{ab}\left(\phi\left({\bf R}+\frac{{\bf r}}{2},{\bf R};t\right)T^{A}\phi\left({\bf R},{\bf R}-\frac{{\bf r}}{2};t\right)\right)_{b'a'} \\
 & +\frac{1}{N_{c}}\phi_{ab}\left({\bf R}-\frac{{\bf r}}{2},{\bf R}+\frac{{\bf r}}{2};t\right)\phi_{b'a'}\left({\bf R}+\frac{{\bf r}}{2},{\bf R}-\frac{{\bf r}}{2};t\right)\,.
 \end{split}
\end{equation}

Using this, we get that

\begin{equation}
\begin{split}
S({\bf R},{\bf r},t)&=W_{SS}({\bf R},{\bf r};t,t_{0})S({\bf R},{\bf r},t_{0})\\
&+W_{SO}^{A}({\bf R},{\bf r};t,t_{0})O^{A}({\bf R},{\bf r},t_{0})\,,
\end{split}
\label{eq:Sevo}
\end{equation}

and

\begin{equation}
\begin{split}
O^{A}({\bf R},{\bf r},t)&=W_{OO}^{AB}({\bf R},{\bf r};t,t_{0})O^{B}({\bf R},{\bf r},t_{0})\\
&+W_{OS}^{A}({\bf R},{\bf r};t,t_{0})S({\bf R},{\bf r},t_{0})\,,
\end{split}
\label{eq:Oevo}
\end{equation}

where

\begin{equation}
\begin{split}
W_{SS}({\bf R},{\bf r};t,t_{0})&=\frac{1}{N_{c}}Tr\left(U^{\dagger}\left(t,t_{0};{\bf R}-\frac{{\bf r}}{2}\right)\phi\left({\bf R}-\frac{{\bf r}}{2},{\bf R}+\frac{{\bf r}}{2};t\right)U\left(t,t_{0};{\bf R}+\frac{{\bf r}}{2}\right)\times\right.\\
&\left.\times\phi\left({\bf R}+\frac{{\bf r}}{2},{\bf R}-\frac{{\bf r}}{2};t_{0}\right)\right)\,,
\end{split}
\end{equation}

\begin{equation}
\begin{split}
W_{SO}^{A}({\bf R},{\bf r};t,t_{0}) & = \sqrt{\frac{2}{N_{c}}}Tr\left(U^{\dagger}\left(t,t_{0};{\bf R}-\frac{{\bf r}}{2}\right)\phi\left({\bf R}-\frac{{\bf r}}{2},{\bf R}+\frac{{\bf r}}{2};t\right)U\left(t,t_{0};{\bf R}+\frac{{\bf r}}{2}\right)\times\right.\\
&\left.\times\phi\left({\bf R}+\frac{{\bf r}}{2},{\bf R};t_{0}\right)T^{A}\phi\left({\bf R},{\bf R}-\frac{{\bf r}}{2};t_{0}\right)\right)\,,
\end{split}
\end{equation}

\begin{equation}
\begin{split}
W_{OO}^{AB}({\bf R},{\bf r};t,t_{0}) & =2Tr\left(U^{\dagger}\left(t,t_{0};{\bf R}-\frac{{\bf r}}{2}\right)\phi\left({\bf R}-\frac{{\bf r}}{2},{\bf R};t\right)T^{A}\phi\left({\bf R},{\bf R}+\frac{{\bf r}}{2};t\right)U\left(t,t_{0};{\bf R}+\frac{{\bf r}}{2}\right)\times\right.\\
&\left.\times\phi\left({\bf R}+\frac{{\bf r}}{2},{\bf R};t_{0}\right)T^{B}\phi\left({\bf R},{\bf R}-\frac{{\bf r}}{2};t_{0}\right)\right)\,,
\end{split}
\end{equation}

and

\begin{equation}
\begin{split}
W_{OS}^{A}({\bf R},{\bf r};t,t_{0})&=\sqrt{\frac{2}{N_{c}}}Tr\left(U^{\dagger}\left(t,t_{0};{\bf R}-\frac{{\bf r}}{2}\right)\phi\left({\bf R}-\frac{{\bf r}}{2},{\bf R};t\right)T^{A}\phi\left({\bf R},{\bf R}+\frac{{\bf r}}{2};t\right)\times\right.\\
&\left.\times U\left(t,t_{0};{\bf R}+\frac{{\bf r}}{2}\right)\phi\left({\bf R}+\frac{{\bf r}}{2},{\bf R}-\frac{{\bf r}}{2};t_{0}\right)\right)\,.
\end{split}
\end{equation}
These equations are more intuitively written in the graphical notation introduced in the appendix \ref{sec:birdtrack}. In fig. \ref{fig:at0tos}, we show the equation that corresponds to the singlet while in fig. \ref{fig:at0too}, we do the same for the octet.

\begin{figure}
\beginpgfgraphicnamed{fig1}
\begin{tikzpicture}
\path (0,0) node {$S(t)=\frac{1}{\sqrt{N_c}}$}
 (1.25,1) node [shape=circle,draw] {$\psi$}
 (1.25,-1) node [shape=circle,draw] {$\chi^\dagger$}
 (1.25,-2) node {$t$}
 (5,1) node [shape=circle,draw] {$\psi$}
 (5,-1) node [shape=circle,draw] {$\chi^\dagger$}
 (5,-2) node {$t_0$}
 (3,-2) node {$t$}
 (6.1,-2) node {$t$}
 (9.1,-2) node {$t_0$}
 (11.6,-2) node {$t$}
 (14.6,-2) node {$t_0$}
 (5,0) node [anchor=west] {$=\frac{1}{N_c}$}
 (15.5,0) node [shape=rectangle,draw] {$O(t_0)$};
\draw [dashed] (14.6,0)--(15,0);
\begin{scope}[very thick]
\draw[->] (1.25,0.6)--(1.25,0) node [anchor=west] {$\,\,=\frac{1}{\sqrt{N_c}}$};
\draw (1.25,0)--(1.25,-0.6);
\draw[->] (3,1)--(3,0);
\draw (4.6,1)--(3,1)--(3,-1)--(4.6,-1);
\draw (6.1,-1)--(6.1,1)--(9.1,1)--(9.1,-1)--(6.1,-1);
\draw[->] (6.1,1)--(6.1,0);
\draw[->] (6.1,-1)--(7.6,-1);
\draw[->] (9.1,-1)--(9.1,0) node [anchor=west] {$S(t_0)+\sqrt{\frac{2}{N_c}}$};
\draw[->] (9.1,1)--(7.6,1);
\draw (11.6,-1)--(11.6,1)--(14.6,1)--(14.6,-1)--(11.6,-1);
\draw[->] (11.6,1)--(11.6,0);
\draw[->] (11.6,-1)--(13.1,-1);
\draw[->] (14.6,-1)--(14.6,0);
\draw[->] (14.6,1)--(13.1,1);
\end{scope}
\end{tikzpicture}
\endpgfgraphicnamed
\caption{The singlet field at time $t$ written as a function of quarkonium fields at time $t_0$. In these diagrams, time goes from right to left so that fields on the right are at time $t_0$ while on the left they are at time $t$.}
\label{fig:at0tos}
\end{figure}
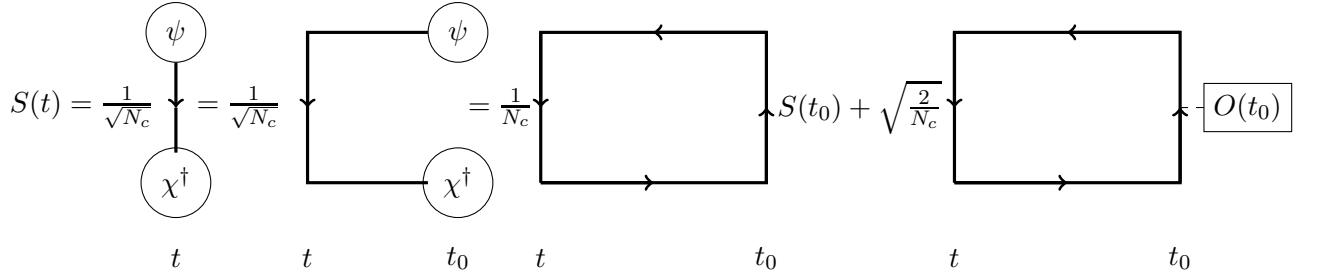
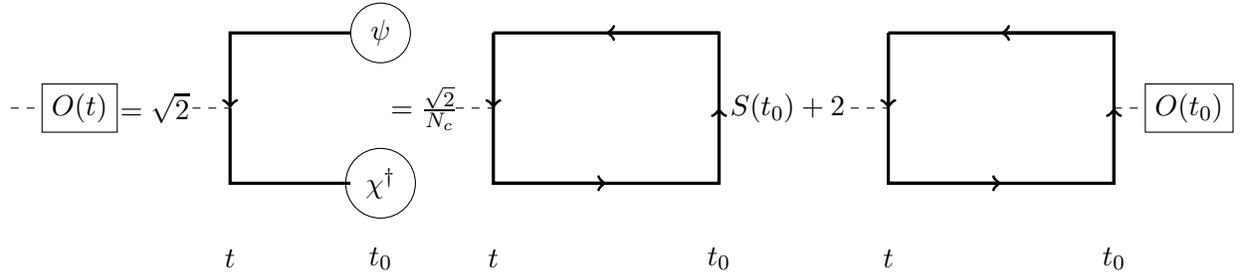
\begin{figure}
\beginpgfgraphicnamed{fig2}
\begin{tikzpicture}
\path (0,0) node [shape=rectangle,draw] {$O(t)$};
\path (1,0) node {$=\sqrt{2}$};
\path (4,1) node [shape=circle,draw] {$\psi$}
 (4,-1) node [shape=circle,draw] {$\chi^\dagger$}
 (4,-2) node {$t_0$}
 (2,-2) node {$t$}
 (5.5,-2) node {$t$}
 (8.5,-2) node {$t_0$}
 (13.75,-2) node {$t_0$}
 (10.75,-2) node {$t$}
 (4,0) node [anchor=west] {$=\frac{\sqrt{2}}{N_c}$};
\path (14.75,0) node [shape=rectangle,draw] {$O(t_0)$};
\draw [dashed] (-0.6,0)--(-1,0);
\draw [dashed] (1.5,0)--(2,0);
\draw [dashed] (5,0)--(5.5,0);
\draw [dashed] (10.25,0)--(10.75,0);
\draw [dashed] (13.75,0)--(14.1,0);
\begin{scope}[very thick]
\draw[->] (3.6,1)--(2,1)--(2,0);
\draw (2,0)--(2,-1)--(3.6,-1);
\draw (5.5,1)--(5.5,-1)--(8.5,-1)--(8.5,1)--(5.5,1);
\draw[->] (5.5,1)--(5.5,0);
\draw[->] (5.5,-1)--(7,-1);
\draw[->] (8.5,-1)--(8.5,0) node [anchor=west] {$S(t_0)+2$};
\draw[->] (8.5,1)--(7,1);
\draw (10.75,1)--(10.75,-1)--(13.75,-1)--(13.75,1)--(10.75,1);
\draw[->] (10.75,1)--(10.75,0);
\draw[->] (10.75,-1)--(12.25,-1);
\draw[->] (13.75,-1)--(13.75,0);
\draw[->] (13.75,1)--(12.25,1);
\end{scope}
\end{tikzpicture}
\endpgfgraphicnamed
\caption{Same as fig. \ref{fig:at0tos} but for the octet.}
\label{fig:at0too}
\end{figure}
Now let us introduce a state which contains a static quark pair in
a singlet state and an arbitrary configuration of light quarks and
gluons (such that Gauss's law is fulfilled)
\begin{equation}
\begin{split}
\tilde{\rho}_{s}=\\
\int\,d^{3}R\,d^{3}r\,d^{3}R'\,d^{3}r'D_{s}({\bf R},{\bf r},{\bf R}',{\bf r}';t)S^{\dagger}({\bf R},{\bf r},t)\rho_{l}S({\bf R}',{\bf r}',t)\,,
\end{split}
\end{equation}
where $\rho_{l}$ represents the state of light quarks and gluons.
The meaning of $D_{s}$ is a generalization of the wave function we
used in eq. (\ref{eq:ws})

\begin{equation}
P_{s}({\bf R},{\bf r},t|\tilde{\rho}_{s})=Tr(S^{\dagger}({\bf R},{\bf r},t)S({\bf R},{\bf r},t)\tilde{\rho}_{s})=D_{s}({\bf R},{\bf r},{\bf R},{\bf r};t)\,.
\label{eq:Pstos}
\end{equation}

In this last equation, we have introduced a new notation that we are
going to use from now on. $P_{s}({\bf R},{\bf r},t|\rho)$ is the conditional
probability to find a singlet at a given time and position provided
that the system is in a state $\rho$. From now on, we define $\rho_s$ (without the tilde) to be a state such that $D_o(t_0)=0$ and $\rho_o$ a state such that $D_s(t_0)=0$. In other words, we consider states for which we know that at a given time, the heavy quarks are in a singlet (octet) state. 

We can make a similar construction for the octet

\begin{equation}
\begin{split}
\tilde{\rho}_{o} &=\\
&\frac{1}{N_{c}^{2}-1}\int\,d^{3}R\,d^{3}r\,d^{3}R'\,d^{3}r'D_{o}({\bf R},{\bf r},{\bf R}',{\bf r}';t) W^{AB}\left(\frac{{\bf R}+{\bf R}'}{2},{\bf R};t\right)O^{B\dagger}({\bf R},{\bf r},t)\tilde{\rho_{l}}\times \\
&\times O^{C}({\bf R}',{\bf r}',t)W^{AC\dagger}\left(\frac{{\bf R}+{\bf R}'}{2},{\bf R}';t\right)\,,
\end{split}
\label{eq:rhootilde}
\end{equation}

Then 
\begin{equation}
P_{o}({\bf R},{\bf r},t|\tilde{\rho}_{o})=Tr(O^{A\dagger}({\bf R},{\bf r},t)O^{A}({\bf R},{\bf r},t)\tilde{\rho}_{o})=D_{o}({\bf R},{\bf r},{\bf R},{\bf r};t)\,.
\end{equation}
Note that in eq. (\ref{eq:rhootilde}) we have introduced a new matrix called $\tilde{\rho}_l$. This is a state of light quarks and gluons such that $\tilde{\rho}_o$ fulfills Gauss's law \footnote{We would like to note that a density matrix as the one discussed below eq. (\ref{eq:Po}), that can be written as a linear combination of pure state density matrices of states as the ones defined in eqs. (\ref{eq:ws}) and (\ref{eq:wo}), can also be written as a linear combination of $\tilde{\rho}_s$ and $\tilde{\rho}_o$}. An approximation that is usually implicit when doing perturbative computations of quarkonium in a medium is $\tilde{\rho}_l=\rho_l$. Imagine that we start with a generic $\rho_l$, and we construct $\tilde{\rho}_l$ by adding to $\rho_l$ an arbitrary gluonic configuration such that $\tilde{\rho}_o$ fulfills Gauss's law. Then, we can imagine that if the environment is big enough $\mathit{Tr}(A\rho_l)\sim\mathit{Tr}(A\tilde{\rho}_l)$ for any operator $A$.  However, this might be problematic for non-perturbative computations. Therefore, we mark in our formulas the difference between $\rho_l$ and $\tilde{\rho}_l$. In fact, the observation that $\tilde{\rho}_l\neq\rho_l$ in general is related with the fact that it is difficult to compute the octet potential in lattice QCD \cite{Philipsen:2013ysa,Bala:2020tdt}. Similar complications were pointed out for the octet free energy in \cite{Jahn:2004qr}. However, a gauge invariant definition of the octet free energy proposed in \cite{Brambilla:2010xn} has been recently computed in lattice QCD \cite{Bazavov:2018wmo}.

\begin{figure}
\beginpgfgraphicnamed{fig3}
\begin{tikzpicture}
\path (0,0) node {$\frac{1}{N_c^2}\mathit{Tr}$};
\begin{scope}[very thick]
\draw (1,1.5) .. controls (0.5,1) and (0.5,-1) .. (1,-1.5);   
\draw (10,1.5) .. controls (10.5,1) and (10.5,-1) .. (10,-1.5);
\draw[->] (1.2,-1)--(3.2,-1);
\draw (3.2,-1)--(5.2,-1) node[anchor=north] {$t$};   
\draw[->] (5.2,-1)--(5.2,0); 
\draw (5.2,0)--(5.2,1);
\draw[->] (5.2,1)--(3.2,1);
\draw (3.2,1)--(1.2,1);  
\draw[->] (1.2,1)--(1.2,0);
\draw (1.2,0)--(1.2,-1) node[anchor=north] {$t_0$};
\draw[->] (5.6,-1)--(7.6,-1);
\draw (7.6,-1)--(9.6,-1) node[anchor=north] {$t_0$};   
\draw[->] (9.6,-1)--(9.6,0) node[anchor=west] {$\,\rho_l$}; 
\draw (9.6,0)--(9.6,1);
\draw[->] (9.6,1)--(7.6,1);
\draw (7.6,1)--(5.6,1);  
\draw[->] (5.6,1)--(5.6,0); 
\draw (5.6,0)--(5.6,-1) node[anchor=north] {$t$};
\end{scope}
\end{tikzpicture}
\endpgfgraphicnamed
\caption{Graphical representation of the transition of a singlet at time $t_0$ to a singlet at time $t$.}
\label{eq:Wlss}
\end{figure}
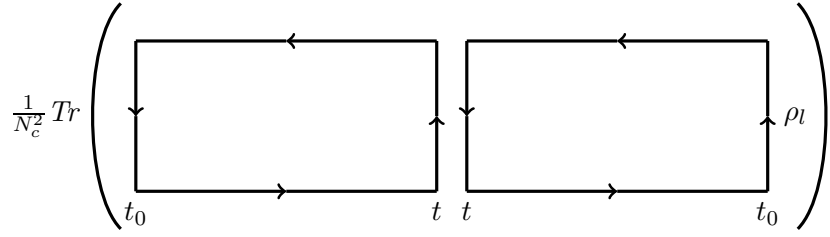

To compute the conditional probability to find a singlet state at time
$t>t_{0}$ provided that we know that at time $t_{0}$ the pair is also in a singlet state, we apply eq. (\ref{eq:Sevo}) to get 

\begin{equation}
P_{s}({\bf R},{\bf r},t|\rho_{s})=Tr(W_{SS}^{\dagger}({\bf R},{\bf r};t,t_{0})W_{SS}({\bf R},{\bf r};t,t_{0})\rho_{l})D_{s}({\bf R},{\bf r},{\bf R},{\bf r};t_{0})\,.
\end{equation}
We find it convenient to represent this equation graphically as shown in fig. \ref{eq:Wlss}

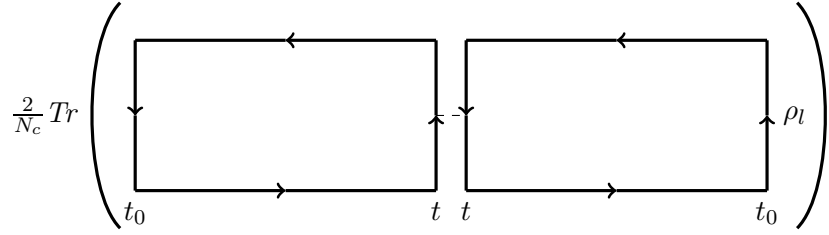
\begin{figure}
\beginpgfgraphicnamed{fig4}
\begin{tikzpicture}
\path (0,0) node {$\frac{2}{N_c}\mathit{Tr}$};
\begin{scope}[very thick]
\draw (1,1.5) .. controls (0.5,1) and (0.5,-1) .. (1,-1.5);   
\draw (10,1.5) .. controls (10.5,1) and (10.5,-1) .. (10,-1.5);
\draw[->] (1.2,-1)--(3.2,-1);
\draw (3.2,-1)--(5.2,-1) node[anchor=north] {$t$};   
\draw[->] (5.2,-1)--(5.2,0); 
\draw (5.2,0)--(5.2,1);
\draw[->] (5.2,1)--(3.2,1);
\draw (3.2,1)--(1.2,1);  
\draw[->] (1.2,1)--(1.2,0);
\draw (1.2,0)--(1.2,-1) node[anchor=north] {$t_0$};
\draw[->] (5.6,-1)--(7.6,-1);
\draw (7.6,-1)--(9.6,-1) node[anchor=north] {$t_0$};   
\draw[->] (9.6,-1)--(9.6,0) node[anchor=west] {$\,\rho_l$}; 
\draw (9.6,0)--(9.6,1);
\draw[->] (9.6,1)--(7.6,1);
\draw (7.6,1)--(5.6,1);  
\draw[->] (5.6,1)--(5.6,0); 
\draw (5.6,0)--(5.6,-1) node[anchor=north] {$t$};
\end{scope}
\draw [dashed] (5.2,0)--(5.6,0);
\end{tikzpicture}
\endpgfgraphicnamed
\caption{Graphical representation of the transition of a singlet at time $t_0$ to an octet at time $t$.}
\label{eq:Wlso}
\end{figure}

Analogously, we apply eq. (\ref{eq:Oevo}) to get

\begin{equation}
P_{o}({\bf R},{\bf r},t|\rho_{s})=Tr(W_{OS}^{A\dagger}({\bf R},{\bf r};t,t_{0})W_{OS}^{A}({\bf R},{\bf r};t,t_{0})\rho_{l})D_{s}({\bf R},{\bf r},{\bf R},{\bf r};t_{0})\,,
\end{equation}

which can be represented graphically as shown in fig. \ref{eq:Wlso}. 

\begin{figure}
\beginpgfgraphicnamed{fig5}
\begin{tikzpicture}
\path (-1.5,0) node {$\frac{2}{N_c(N_c^2-1)}\mathit{Tr}$};
\path (11.5,0) node [anchor=east] {$\tilde{\rho}_l\,$};
\draw [dashed] (1.2,0)--(0.5,0)--(0.5,1.5)--(10.5,1.5)--(10.5,0)--(9.6,0);
\begin{scope}[very thick]
\draw (0,1.5) .. controls (-0.5,1) and (-0.5,-1) .. (0,-1.5);   
\draw (11,1.5) .. controls (11.5,1) and (11.5,-1) .. (11,-1.5);
\draw[->] (1.2,-1)--(3.2,-1);
\draw (3.2,-1)--(5.2,-1) node[anchor=north] {$t$};   
\draw[->] (5.2,-1)--(5.2,0); 
\draw (5.2,0)--(5.2,1);
\draw[->] (5.2,1)--(3.2,1);
\draw (3.2,1)--(1.2,1);  
\draw[->] (1.2,1)--(1.2,0);
\draw (1.2,0)--(1.2,-1) node[anchor=north] {$t_0$};
\draw[->] (5.6,-1)--(7.6,-1);
\draw (7.6,-1)--(9.6,-1) node[anchor=north] {$t_0$};   
\draw[->] (9.6,-1)--(9.6,0); 
\draw (9.6,0)--(9.6,1);
\draw[->] (9.6,1)--(7.6,1);
\draw (7.6,1)--(5.6,1);  
\draw[->] (5.6,1)--(5.6,0); 
\draw (5.6,0)--(5.6,-1) node[anchor=north] {$t$};
\end{scope}
\end{tikzpicture}
\endpgfgraphicnamed
\caption{Graphical representation of the transition of an octet at time $t_0$ to a singlet at time $t$.}
\label{eq:Wlos}
\end{figure}
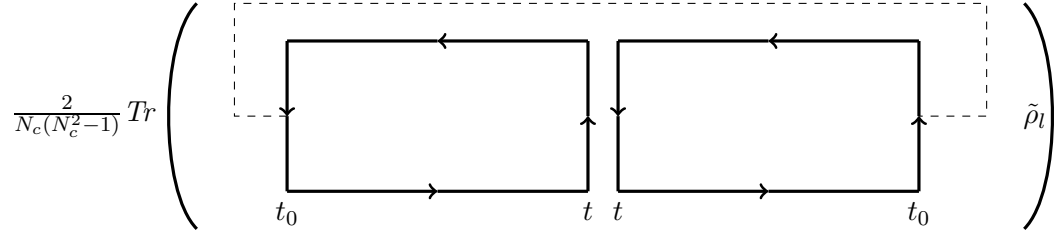

Equivalently, we get the following results for the case in which we
know that the pair is in an octet state at $t_{0}$. The conditional
probability of having a singlet at time $t$ is 

\begin{equation}
P_{s}({\bf R},{\bf r},t|\rho_{o})=\frac{1}{N_{c}^{2}-1}Tr(W_{SO}^{A\dagger}({\bf R},{\bf r};t,t_{0})W_{SO}^{A}({\bf R},{\bf r};t,t_{0})\tilde{\rho}_{l})D_{o}({\bf R},{\bf r},{\bf R},{\bf r};t_{0})\,,
\end{equation}

which we represent graphically in fig. \ref{eq:Wlos}.

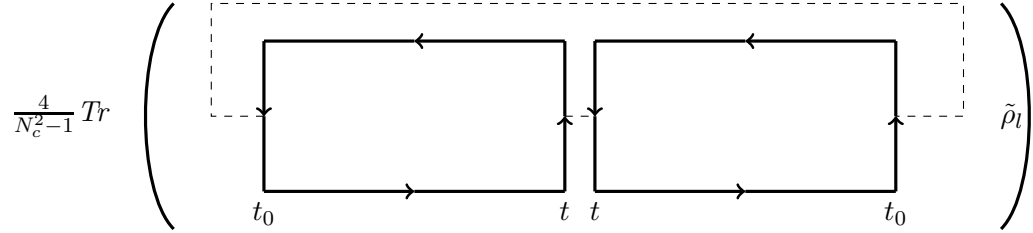
\begin{figure}
\beginpgfgraphicnamed{fig6}
\begin{tikzpicture}
\path (-1.5,0) node {$\frac{4}{N_c^2-1}\mathit{Tr}$};
\path (11.5,0) node [anchor=east] {$\tilde{\rho}_l\,$};
\draw [dashed] (1.2,0)--(0.5,0)--(0.5,1.5)--(10.5,1.5)--(10.5,0)--(9.6,0);
\draw [dashed] (5.2,0)--(5.6,0);
\begin{scope}[very thick]
\draw (0,1.5) .. controls (-0.5,1) and (-0.5,-1) .. (0,-1.5);   
\draw (11,1.5) .. controls (11.5,1) and (11.5,-1) .. (11,-1.5);
\draw[->] (1.2,-1)--(3.2,-1);
\draw (3.2,-1)--(5.2,-1) node[anchor=north] {$t$};   
\draw[->] (5.2,-1)--(5.2,0); 
\draw (5.2,0)--(5.2,1);
\draw[->] (5.2,1)--(3.2,1);
\draw (3.2,1)--(1.2,1);  
\draw[->] (1.2,1)--(1.2,0);
\draw (1.2,0)--(1.2,-1) node[anchor=north] {$t_0$};
\draw[->] (5.6,-1)--(7.6,-1);
\draw (7.6,-1)--(9.6,-1) node[anchor=north] {$t_0$};   
\draw[->] (9.6,-1)--(9.6,0); 
\draw (9.6,0)--(9.6,1);
\draw[->] (9.6,1)--(7.6,1);
\draw (7.6,1)--(5.6,1);  
\draw[->] (5.6,1)--(5.6,0); 
\draw (5.6,0)--(5.6,-1) node[anchor=north] {$t$};
\end{scope}
\end{tikzpicture}
\endpgfgraphicnamed
\caption{Graphical representation of the transition of an octet at time $t_0$ to an octet at time $t$.}
\label{eq:Wloo}
\end{figure}

For the probability to find an octet we have 

\begin{equation}
P_{o}({\bf R},{\bf r},t|\rho_{o})=\frac{1}{N_{c}^{2}-1}Tr(W_{OO}^{AB\dagger}({\bf R},{\bf r};t,t_{0})W_{OO}^{AB}({\bf R},{\bf r};t,t_{0})\tilde{\rho}_{l})D_{o}({\bf R},{\bf r},{\bf R},{\bf r};t_{0})\,,
\end{equation}

which we represent graphically in fig. \ref{eq:Wloo}.

We focused on the study of conditional probabilities in this section. It is straightforward to generalize these results to the study of $D_{s,o}(\mathbf{R},\mathbf{r},\mathbf{R}',\mathbf{r}';t)$ given that we know that the system was in a singlet (or octet) state at time $t_0<t$. However, this would have made the notation more cumbersome. Nevertheless, we do present results for $\mathbf{r}\neq\mathbf{r}'$ in section \ref{sec:class} and in appendix \ref{sec:expblocks}.
\section{Connection with the physics of heavy quarkonium}
\label{sec:conn}
In the rest of the sections of this paper, we focus on studying the static limit ($m\to\infty$). However, in this section, we argue that our results are also useful to describe bound states of heavy quarks with finite mass. To illustrate this let us now discuss the situation at $T=0$ in the non-relativistic EFT framework. Non-relativistic EFTs exploit the separation between the scales relevant for quarkonium physics. They are the mass of heavy quarkonium $m$, the inverse of the typical radius $\frac{1}{r}$ (which scales like $mv$, with $v\ll 1$) and the binding energy $E$ (of order $mv^2$).

pNRQCD is the EFT that is obtained after integrating out the scales $m$ and $mv$. The Lagrangian of pNRQCD can be ordered as an infinite sum of operators multiplied by Wilson coefficients with different powers of $m$ and $r$. However, given a desired precision, there exists a power counting that ensures that only a finite number of operators needs to be taken into account. The heavy quarkonium potential is a Wilson coefficient of pNRQCD which needs to be taken into account at leading order. 

The heavy quarkonium potential of pNRQCD is indeed equal to the static potential. This can be shown by noting that the matching to obtain the Wilson coefficients can be done in the limit $m\to\infty$. Once this is done the matched potential can be used in any computation with a finite value of $m$. More generally, there is a well defined method that allows one to connect expectation values of correlators of the interpolating fields with expectation values of pNRQCD singlet and octet fields. More details on this procedure can be found in \cite{Brambilla:2004jw}.

We will try to give here a more physical and intuitive idea of why this is possible. Static quarkonium is characterized by the fact that the positions of the heavy quark and the antiquark do not change in time. In the case of real quarkonium, the position of the heavy quarks will look almost static if we are interested in what happens at time scales much smaller than $\frac{1}{mv^2}$. The heavy quarkonium potential is obtained after integrating out the scale $\frac{1}{r}$. Two types of particles are important in this regime:
\begin{itemize}
    \item Soft particles, with momentum and energy scaling like $mv$.
    \item Potential particles, with momentum of order $mv$ and energy of order $mv^2$.
\end{itemize}
From the point of view of soft particles, the heavy quarks are essentially frozen. Regarding potential particles, in most gauges, they are related to instantaneous interactions. The reason is that we can approximate their propagator with the limit of zero energy. The conclusion is that, for the physics that is encoded in the Wilson coefficients, the heavy quark and the antiquark do look like static particles. 

Now let us look at how we could generalize this result to the case $\frac{1}{r}\sim T$, where $T$ is the temperature. Note that the case $\frac{1}{r}\gg T$ was already studied in \cite{Brambilla:2016wgg,Brambilla:2017zei}. A rigorous way to do this would be to define a version of pNRQCD valid in this situation. However the effects of the scale $\frac{1}{r}\sim T$ cannot be encoded in an effective action because of the existence of dissipative effects. We would need a different type of EFT in which we consider operators that mix fields in different branches of the Schwinger-Keldysh contour (examples of this type of EFT are found when studying hydrodynamics, see \cite{Jain:2020vgc} and references therein). 

This version of pNRQCD is still unknown, and its construction is beyond the scope of this work. However, we can still use the intuitive and physical arguments related to the time scales. In order for a non-relativistic bound state to exist, we need the time scale at which the effective interaction generated by the physics at the scale $\frac{1}{r}\sim T$ becomes important to be much larger than $\frac{1}{r}$. In this case, in order to compute the effective interaction originating from the scale $\frac{1}{r}\sim T$ and by the same arguments that are valid in the $T=0$ case, approximating real quarkonium by a static pair of heavy quarks is valid.

Finally, let us warn again that we made an abuse of notation naming $S$ (singlet) and $O^A$ (octet) for the fields previously defined. More rigorously, they are interpolating fields of real pNRQCD singlets and octets (see \cite{Brambilla:2004jw}). The connection between these two kinds of fields is something that needs to be worked out. However, if the intuition from the $T=0$ case is valid, we can perform the matching at leading order ignoring the difference between the singlet or octet field and the corresponding interpolating field. Corrections would be of size $rE$ ($E$ being the binding energy). The arbitrariness on the way we chose the space-like Wilson lines that make the fields $S$ and $O^A$ transform in the desired way is related to this issue. 
\section{Use of the Fierz identity and the large $N_{c}$ limit\label{sec:Results-in-the}}
\label{sec:largenc}
\begin{figure}
\beginpgfgraphicnamed{fig7a}
\begin{tikzpicture}
\draw [gluon] (3,1)--(5,-1);
\path (1.2,0) node [anchor=east] {$\frac{g^2}{N_c^2}$}
(9.6,0) node [anchor=west] {$\propto 1$};
\begin{scope}[very thick]
\draw[->] (1.2,-1)--(3.2,-1);
\draw (3.2,-1)--(5.2,-1) node[anchor=north] {$t$};   
\draw[->] (5.2,-1)--(5.2,0); 
\draw (5.2,0)--(5.2,1);
\draw[->] (5.2,1)--(3.2,1);
\draw (3.2,1)--(1.2,1);  
\draw[->] (1.2,1)--(1.2,0);
\draw (1.2,0)--(1.2,-1) node[anchor=north] {$t_0$};
\draw[->] (5.6,-1)--(7.6,-1);
\draw (7.6,-1)--(9.6,-1) node[anchor=north] {$t_0$};   
\draw[->] (9.6,-1)--(9.6,0); 
\draw (9.6,0)--(9.6,1);
\draw[->] (9.6,1)--(7.6,1);
\draw (7.6,1)--(5.6,1);  
\draw[->] (5.6,1)--(5.6,0); 
\draw (5.6,0)--(5.6,-1) node[anchor=north] {$t$};
\end{scope}
\end{tikzpicture}
\endpgfgraphicnamed
\beginpgfgraphicnamed{fig7b}
\begin{tikzpicture}
\draw [gluon] (3.2,1).. controls (5.4,1.5) .. (7.6,1);
\path (1.2,0) node [anchor=east] {$\frac{g^2}{N_c^2}$}
(9.6,0) node [anchor=west] {$\propto \frac{1}{N_c^2}$};
\begin{scope}[very thick]
\draw[->] (1.2,-1)--(3.2,-1);
\draw (3.2,-1)--(5.2,-1) node[anchor=north] {$t$};   
\draw[->] (5.2,-1)--(5.2,0); 
\draw (5.2,0)--(5.2,1);
\draw[->] (5.2,1)--(3.2,1);
\draw (3.2,1)--(1.2,1);  
\draw[->] (1.2,1)--(1.2,0);
\draw (1.2,0)--(1.2,-1) node[anchor=north] {$t_0$};
\draw[->] (5.6,-1)--(7.6,-1);
\draw (7.6,-1)--(9.6,-1) node[anchor=north] {$t_0$};   
\draw[->] (9.6,-1)--(9.6,0); 
\draw (9.6,0)--(9.6,1);
\draw[->] (9.6,1)--(7.6,1);
\draw (7.6,1)--(5.6,1);  
\draw[->] (5.6,1)--(5.6,0); 
\draw (5.6,0)--(5.6,-1) node[anchor=north] {$t$};
\end{scope}
\end{tikzpicture}
\endpgfgraphicnamed
\caption{Illustration of the power counting in $N_c$ of two different types of diagrams at leading order in the singlet to singlet evolution. In contrast to other diagrams in this manuscript, in this case, plane lines represent Kronecker deltas.}
\label{fig:lncWss}
\end{figure}
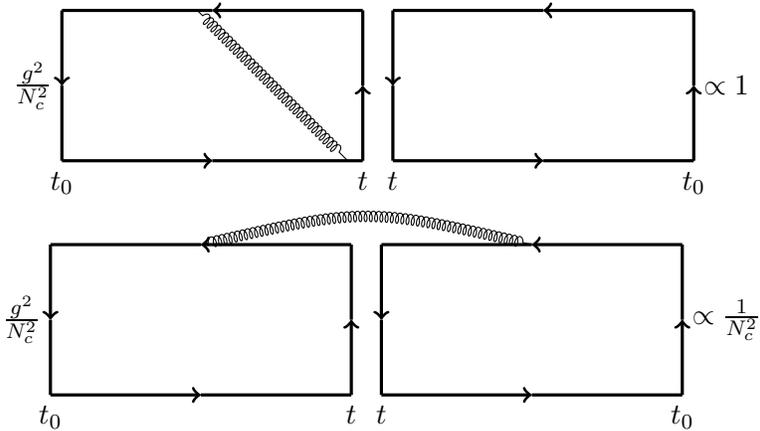

First, we discuss fig. \ref{eq:Wlss} in the large $N_{c}$ limit
with $g^{2}N_{c}\sim 1$ \cite{tHooft:1973alw}.
It scales like $1$ in terms of $N_{c}$. However,
any diagram in which a gluon connects the two Wilson loops will be
suppressed by a factor $\frac{1}{N_{c}^{2}}$. In terms of the Schwinger-Keldysh
contour, this means that eq. (\ref{eq:Pstos}) is dominated by diagrams
which do not mix the two branches of the contour. To illustrate this, we show several examples in fig. \ref{fig:lncWss}. In these diagrams, we change our notation slightly. Plane lines represent a Kronecker delta instead of a Wilson line and we show gluon insertions explicitly.

 This implies that 

\begin{equation}
Tr(W_{SS}^{\dagger}({\bf R},{\bf r};t,t_{0})W_{SS}({\bf R},{\bf r};t,t_{0})\rho_{l})=\left|Tr(W_{SS}({\bf R},{\bf r};t,t_{0})\rho_{l})\right|^{2}+\mathcal{O}\left(\frac{1}{N_{c}^{2}}\right)\,.\label{eq:effh}
\end{equation}

These results suggest that, in this approximation, the survival probability
of a singlet can be encoded in an effective Hamiltonian. $Tr(W_{SS}({\bf R},{\bf r};t,t_{0})\rho_{l})$ is the Wilson loop from which the static potential is obtained in lattice QCD computations \cite{Burnier:2016mxc}.

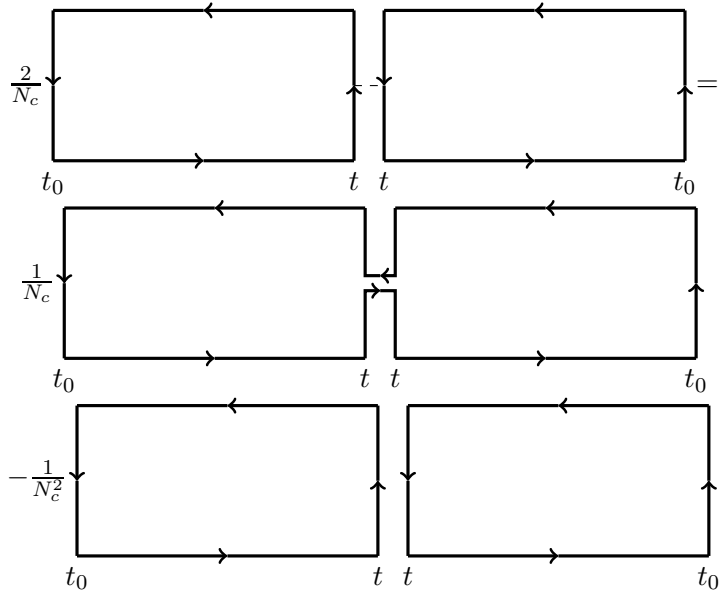
\begin{figure}
\beginpgfgraphicnamed{fig8a}
\begin{tikzpicture}
\path (1.2,0) node [anchor=east] {$\frac{2}{N_c}$};
\begin{scope}[very thick]
\draw[->] (1.2,-1)--(3.2,-1);
\draw (3.2,-1)--(5.2,-1) node[anchor=north] {$t$};   
\draw[->] (5.2,-1)--(5.2,0); 
\draw (5.2,0)--(5.2,1);
\draw[->] (5.2,1)--(3.2,1);
\draw (3.2,1)--(1.2,1);  
\draw[->] (1.2,1)--(1.2,0);
\draw (1.2,0)--(1.2,-1) node[anchor=north] {$t_0$};
\draw[->] (5.6,-1)--(7.6,-1);
\draw (7.6,-1)--(9.6,-1) node[anchor=north] {$t_0$};   
\draw[->] (9.6,-1)--(9.6,0) node[anchor=west] {$=$}; 
\draw (9.6,0)--(9.6,1);
\draw[->] (9.6,1)--(7.6,1);
\draw (7.6,1)--(5.6,1);  
\draw[->] (5.6,1)--(5.6,0); 
\draw (5.6,0)--(5.6,-1) node[anchor=north] {$t$};
\end{scope}
\draw [dashed] (5.2,0)--(5.6,0);
\end{tikzpicture}
\endpgfgraphicnamed
\beginpgfgraphicnamed{fig8b}
\begin{tikzpicture}
\path (1.2,0) node [anchor=east] {$\frac{1}{N_c}$};
\begin{scope}[very thick]
\draw[->] (1.2,-1)--(3.2,-1);
\draw (3.2,-1)--(5.2,-1) node[anchor=north] {$t$};   
\draw[->] (5.2,-1)--(5.2,-0.1)--(5.4,-0.1); 
\draw (5.4,0.1)--(5.2,0.1)--(5.2,1);
\draw[->] (5.2,1)--(3.2,1);
\draw (3.2,1)--(1.2,1);  
\draw[->] (1.2,1)--(1.2,0);
\draw (1.2,0)--(1.2,-1) node[anchor=north] {$t_0$};
\draw[->] (5.6,-1)--(7.6,-1);
\draw (7.6,-1)--(9.6,-1) node[anchor=north] {$t_0$};   
\draw[->] (9.6,-1)--(9.6,0); 
\draw (9.6,0)--(9.6,1);
\draw[->] (9.6,1)--(7.6,1);
\draw (7.6,1)--(5.6,1);  
\draw[->] (5.6,1)--(5.6,0.1)--(5.4,0.1); 
\draw (5.4,-0.1)--(5.6,-0.1)--(5.6,-1) node[anchor=north] {$t$};
\end{scope}
\end{tikzpicture}
\endpgfgraphicnamed
\beginpgfgraphicnamed{fig8c}
\begin{tikzpicture}
\path (1.2,0) node [anchor=east] {$-\frac{1}{N_c^2}$};
\begin{scope}[very thick]
\draw[->] (1.2,-1)--(3.2,-1);
\draw (3.2,-1)--(5.2,-1) node[anchor=north] {$t$};   
\draw[->] (5.2,-1)--(5.2,0); 
\draw (5.2,0)--(5.2,1);
\draw[->] (5.2,1)--(3.2,1);
\draw (3.2,1)--(1.2,1);  
\draw[->] (1.2,1)--(1.2,0);
\draw (1.2,0)--(1.2,-1) node[anchor=north] {$t_0$};
\draw[->] (5.6,-1)--(7.6,-1);
\draw (7.6,-1)--(9.6,-1) node[anchor=north] {$t_0$};   
\draw[->] (9.6,-1)--(9.6,0); 
\draw (9.6,0)--(9.6,1);
\draw[->] (9.6,1)--(7.6,1);
\draw (7.6,1)--(5.6,1);  
\draw[->] (5.6,1)--(5.6,0); 
\draw (5.6,0)--(5.6,-1) node[anchor=north] {$t$};
\end{scope}
\end{tikzpicture}
\endpgfgraphicnamed
\caption{Result of applying the Fierz identity to the singlet to octet transition. The diagram in the first line is equivalent to the sum of the diagrams in the second and third lines. All of them are of the same size in the $N_c$ counting.}
\label{fig:lncWso}
\end{figure}

From now on, we
are going to use the Fierz identity, which is represented graphically in Appendix \ref{sec:birdtrack}. For example, using this identity in fig. \ref{eq:Wlos} we get the expansion of fig. \ref{fig:lncWso}. The tree terms appearing in this equation are of order $1$ in the
large $N_{c}$ power counting. The conclusions that we can make up
to now are that neither the survival probability of the singlet nor
the transition from an octet into a singlet are suppressed in the
large $N_{c}$ limit and that the survival probability information
can be encoded in an effective Hamiltonian (meaning that there
is no influence of diagrams connecting the two paths of the Schwinger-Keldysh
contour). We leave to Appendix \ref{sec:expblocks} the writing of the explicit numerical expression.

\begin{figure}
\beginpgfgraphicnamed{fig9a}
\begin{tikzpicture}
\path (0.5,0) node [anchor=east] {$\frac{2}{N_c(N_c^2-1)}$};
\path (11.5,0) node [anchor=east] {$=$};
\draw [dashed] (1.2,0)--(0.5,0)--(0.5,1.5)--(10.5,1.5)--(10.5,0)--(9.6,0);
\begin{scope}[very thick]
\draw[->] (1.2,-1)--(3.2,-1);
\draw (3.2,-1)--(5.2,-1) node[anchor=north] {$t$};   
\draw[->] (5.2,-1)--(5.2,0); 
\draw (5.2,0)--(5.2,1);
\draw[->] (5.2,1)--(3.2,1);
\draw (3.2,1)--(1.2,1);  
\draw[->] (1.2,1)--(1.2,0);
\draw (1.2,0)--(1.2,-1) node[anchor=north] {$t_0$};
\draw[->] (5.6,-1)--(7.6,-1);
\draw (7.6,-1)--(9.6,-1) node[anchor=north] {$t_0$};   
\draw[->] (9.6,-1)--(9.6,0); 
\draw (9.6,0)--(9.6,1);
\draw[->] (9.6,1)--(7.6,1);
\draw (7.6,1)--(5.6,1);  
\draw[->] (5.6,1)--(5.6,0); 
\draw (5.6,0)--(5.6,-1) node[anchor=north] {$t$};
\end{scope}
\end{tikzpicture}
\endpgfgraphicnamed
\vspace{5mm}
\beginpgfgraphicnamed{fig9b}
\begin{tikzpicture}
\path (0.5,0) node [anchor=east] {$\frac{1}{N_c(N_c^2-1)}$};
\begin{scope}[very thick]
\draw[->,gray] (1.2,0.1)--(0.5,0.1)--(0.5,1.5)--(5.5,1.5);
\draw[gray] (5.5,1.5)--(10.5,1.5)--(10.5,0.1)--(9.6,0.1);
\draw[gray] (1.2,-0.1)--(0.5,-0.1)--(0.5,-1.5)--(5.5,-1.5);
\draw[->,gray] (9.6,-0.1)--(10.5,-0.1)--(10.5,-1.5)--(5.5,-1.5);
\draw[->] (1.2,-1)--(3.2,-1);
\draw (3.2,-1)--(5.2,-1) node[anchor=north] {$t$};   
\draw[->] (5.2,-1)--(5.2,0); 
\draw (5.2,0)--(5.2,1);
\draw[->] (5.2,1)--(3.2,1);
\draw (3.2,1)--(1.2,1);  
\draw[->] (1.2,1)--(1.2,0.1);
\draw (1.2,-0.1)--(1.2,-1) node[anchor=north] {$t_0$};
\draw[->] (5.6,-1)--(7.6,-1);
\draw (7.6,-1)--(9.6,-1) node[anchor=north] {$t_0$};   
\draw[->] (9.6,-1)--(9.6,-0.1); 
\draw (9.6,0.1)--(9.6,1);
\draw[->] (9.6,1)--(7.6,1);
\draw (7.6,1)--(5.6,1);  
\draw[->] (5.6,1)--(5.6,0); 
\draw (5.6,0)--(5.6,-1) node[anchor=north] {$t$};
\end{scope}
\end{tikzpicture}
\endpgfgraphicnamed
\vspace{5mm}
\beginpgfgraphicnamed{fig9c}
\begin{tikzpicture}
\path (1.2,0) node [anchor=east] {$-\frac{1}{N_c^2(N_c^2-1)}$};
\path (9.6,0) node[anchor=west] {$=\mathcal{O}\left(\frac{1}{N_c^2}\right)$};
\begin{scope}[very thick]
\draw[->] (1.2,-1)--(3.2,-1);
\draw (3.2,-1)--(5.2,-1) node[anchor=north] {$t$};   
\draw[->] (5.2,-1)--(5.2,0); 
\draw (5.2,0)--(5.2,1);
\draw[->] (5.2,1)--(3.2,1);
\draw (3.2,1)--(1.2,1);  
\draw[->] (1.2,1)--(1.2,0);
\draw (1.2,0)--(1.2,-1) node[anchor=north] {$t_0$};
\draw[->] (5.6,-1)--(7.6,-1);
\draw (7.6,-1)--(9.6,-1) node[anchor=north] {$t_0$};   
\draw[->] (9.6,-1)--(9.6,0); 
\draw (9.6,0)--(9.6,1);
\draw[->] (9.6,1)--(7.6,1);
\draw (7.6,1)--(5.6,1);  
\draw[->] (5.6,1)--(5.6,0); 
\draw (5.6,0)--(5.6,-1) node[anchor=north] {$t$};
\end{scope}
\end{tikzpicture}
\endpgfgraphicnamed
\caption{Result of applying the Fierz identity to the octet to singlet transition. Gray lines indicate Kronecker deltas instead of Wilson lines. In fact, gray lines in this diagram, connect points located at the same time and position but on different branches of the Schwinger-Keldysh contour. Probably there is a more clever way to draw this kind of diagrams along the lines of what was proposed in \cite{Horava:2020apz}, but this is outside the scope of this work and our drawing skills.}
\label{fig:Wlso}
\end{figure}
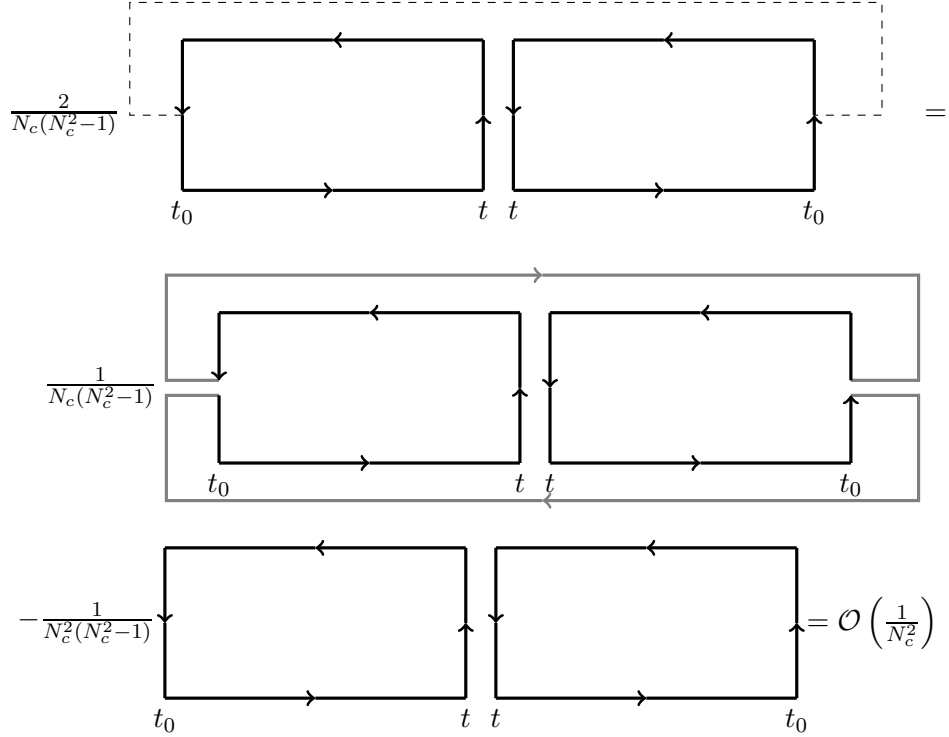

Now we look at fig. \ref{fig:Wlso}. We can see that
in this case both terms are of the same order but suppressed in the
large $N_{c}$ limit which means that the transition from an octet
to a singlet is suppressed if we assume that the populations of singlets and octets are of the same order. This implies that, in the large $N_c$ limit, octet to singlet transitions are not important for computing the number of octets, although it can be relevant for computing the number of singlets (if the number of octets is $N_c^2$ bigger).

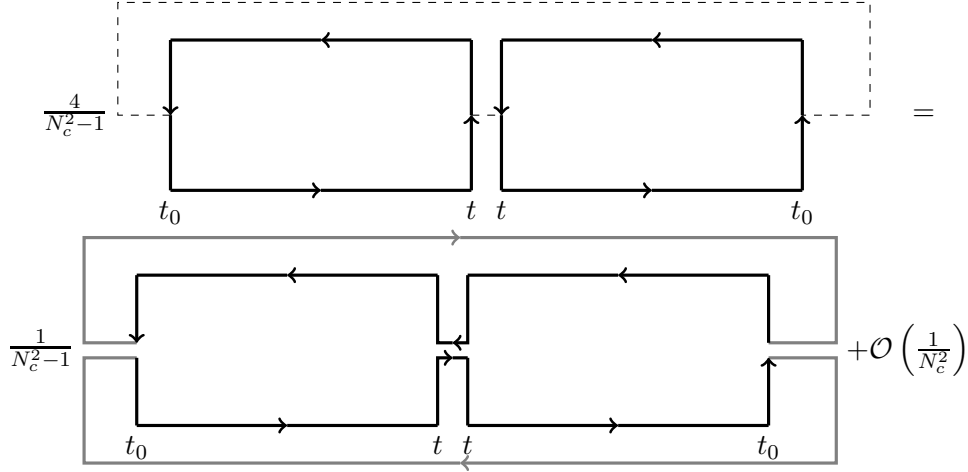
\begin{figure}
\beginpgfgraphicnamed{fig10a}
\begin{tikzpicture}
\path (0.5,0) node [anchor=east] {$\frac{4}{N_c^2-1}$};
\path (11.5,0) node [anchor=east] {$=$};
\draw [dashed] (1.2,0)--(0.5,0)--(0.5,1.5)--(10.5,1.5)--(10.5,0)--(9.6,0);
\draw [dashed] (5.2,0)--(5.6,0);
\begin{scope}[very thick]
\draw[->] (1.2,-1)--(3.2,-1);
\draw (3.2,-1)--(5.2,-1) node[anchor=north] {$t$};   
\draw[->] (5.2,-1)--(5.2,0); 
\draw (5.2,0)--(5.2,1);
\draw[->] (5.2,1)--(3.2,1);
\draw (3.2,1)--(1.2,1);  
\draw[->] (1.2,1)--(1.2,0);
\draw (1.2,0)--(1.2,-1) node[anchor=north] {$t_0$};
\draw[->] (5.6,-1)--(7.6,-1);
\draw (7.6,-1)--(9.6,-1) node[anchor=north] {$t_0$};   
\draw[->] (9.6,-1)--(9.6,0); 
\draw (9.6,0)--(9.6,1);
\draw[->] (9.6,1)--(7.6,1);
\draw (7.6,1)--(5.6,1);  
\draw[->] (5.6,1)--(5.6,0); 
\draw (5.6,0)--(5.6,-1) node[anchor=north] {$t$};
\end{scope}
\end{tikzpicture}
\endpgfgraphicnamed
\beginpgfgraphicnamed{fig10b}
\begin{tikzpicture}
\path (0.5,0) node [anchor=east] {$\frac{1}{N_c^2-1}$};
\path (10.5,0) node [anchor=west] {$+\mathcal{O}\left(\frac{1}{N_c^2}\right)$};
\begin{scope}[very thick]
\draw[->,gray] (1.2,0.1)--(0.5,0.1)--(0.5,1.5)--(5.5,1.5);
\draw[gray] (5.5,1.5)--(10.5,1.5)--(10.5,0.1)--(9.6,0.1);
\draw[gray] (1.2,-0.1)--(0.5,-0.1)--(0.5,-1.5)--(5.5,-1.5);
\draw[->,gray] (9.6,-0.1)--(10.5,-0.1)--(10.5,-1.5)--(5.5,-1.5);
\draw[->] (1.2,-1)--(3.2,-1);
\draw (3.2,-1)--(5.2,-1) node[anchor=north] {$t$};   
\draw[->] (5.2,-1)--(5.2,-0.1)--(5.4,-0.1);
\draw (5.4,-0.1)--(5.6,-0.1); 
\draw (5.4,0.1)--(5.2,0.1)--(5.2,1);
\draw[->] (5.2,1)--(3.2,1);
\draw (3.2,1)--(1.2,1);  
\draw[->] (1.2,1)--(1.2,0.1);
\draw (1.2,-0.1)--(1.2,-1) node[anchor=north] {$t_0$};
\draw[->] (5.6,-1)--(7.6,-1);
\draw (7.6,-1)--(9.6,-1) node[anchor=north] {$t_0$};   
\draw[->] (9.6,-1)--(9.6,-0.1); 
\draw (9.6,0.1)--(9.6,1);
\draw[->] (9.6,1)--(7.6,1);
\draw (7.6,1)--(5.6,1);  
\draw[->] (5.6,1)--(5.6,0.1)--(5.4,0.1); 
\draw (5.6,-0.1)--(5.6,-1) node[anchor=north] {$t$};
\end{scope}
\end{tikzpicture}
\endpgfgraphicnamed
\caption{Result of applying the Fierz identity to the octet to octet transition. We show in the second line the only term that is not suppressed in the large $N_c$ limit. The meaning of gray lines is the same as in fig. \ref{fig:Wlso}}
\label{fig:Wloo}
\end{figure}

\begin{figure}
\beginpgfgraphicnamed{fig11a}
\begin{tikzpicture}
\draw [gluon] (3.2,1).. controls (5.4,1.4) .. (7.6,1);
\path (0.5,0) node [anchor=east] {$\frac{g^2}{N_c^2-1}$};
\path (10.5,0) node [anchor=west] {$\propto 1$};
\begin{scope}[very thick]
\draw[->,gray] (1.2,0.1)--(0.5,0.1)--(0.5,1.5)--(5.5,1.5);
\draw[gray] (5.5,1.5)--(10.5,1.5)--(10.5,0.1)--(9.6,0.1);
\draw[gray] (1.2,-0.1)--(0.5,-0.1)--(0.5,-1.5)--(5.5,-1.5);
\draw[->,gray] (9.6,-0.1)--(10.5,-0.1)--(10.5,-1.5)--(5.5,-1.5);
\draw[->] (1.2,-1)--(3.2,-1);
\draw (3.2,-1)--(5.2,-1) node[anchor=north] {$t$};   
\draw[->] (5.2,-1)--(5.2,-0.1)--(5.4,-0.1);
\draw (5.4,-0.1)--(5.6,-0.1); 
\draw (5.4,0.1)--(5.2,0.1)--(5.2,1);
\draw[->] (5.2,1)--(3.2,1);
\draw (3.2,1)--(1.2,1);  
\draw[->] (1.2,1)--(1.2,0.1);
\draw (1.2,-0.1)--(1.2,-1) node[anchor=north] {$t_0$};
\draw[->] (5.6,-1)--(7.6,-1);
\draw (7.6,-1)--(9.6,-1) node[anchor=north] {$t_0$};   
\draw[->] (9.6,-1)--(9.6,-0.1); 
\draw (9.6,0.1)--(9.6,1);
\draw[->] (9.6,1)--(7.6,1);
\draw (7.6,1)--(5.6,1);  
\draw[->] (5.6,1)--(5.6,0.1)--(5.4,0.1); 
\draw (5.6,-0.1)--(5.6,-1) node[anchor=north] {$t$};
\end{scope}
\end{tikzpicture}
\endpgfgraphicnamed
\beginpgfgraphicnamed{fig11b}
\begin{tikzpicture}
\draw [gluon] (3,1)--(5,-1);
\path (0.5,0) node [anchor=east] {$\frac{g^2}{N_c^2-1}$};
\path (10.5,0) node [anchor=west] {$\propto \frac{1}{N_c^2}$};
\begin{scope}[very thick]
\draw[->,gray] (1.2,0.1)--(0.5,0.1)--(0.5,1.5)--(5.5,1.5);
\draw[gray] (5.5,1.5)--(10.5,1.5)--(10.5,0.1)--(9.6,0.1);
\draw[gray] (1.2,-0.1)--(0.5,-0.1)--(0.5,-1.5)--(5.5,-1.5);
\draw[->,gray] (9.6,-0.1)--(10.5,-0.1)--(10.5,-1.5)--(5.5,-1.5);
\draw[->] (1.2,-1)--(3.2,-1);
\draw (3.2,-1)--(5.2,-1) node[anchor=north] {$t$};   
\draw[->] (5.2,-1)--(5.2,-0.1)--(5.4,-0.1);
\draw (5.4,-0.1)--(5.6,-0.1); 
\draw (5.4,0.1)--(5.2,0.1)--(5.2,1);
\draw[->] (5.2,1)--(3.2,1);
\draw (3.2,1)--(1.2,1);  
\draw[->] (1.2,1)--(1.2,0.1);
\draw (1.2,-0.1)--(1.2,-1) node[anchor=north] {$t_0$};
\draw[->] (5.6,-1)--(7.6,-1);
\draw (7.6,-1)--(9.6,-1) node[anchor=north] {$t_0$};   
\draw[->] (9.6,-1)--(9.6,-0.1); 
\draw (9.6,0.1)--(9.6,1);
\draw[->] (9.6,1)--(7.6,1);
\draw (7.6,1)--(5.6,1);  
\draw[->] (5.6,1)--(5.6,0.1)--(5.4,0.1); 
\draw (5.6,-0.1)--(5.6,-1) node[anchor=north] {$t$};
\end{scope}
\end{tikzpicture}
\endpgfgraphicnamed
    \caption{Illustration of the power counting in $N_c$ of two different types of diagrams at leading order in the octet to octet evolution. In contrast to other diagrams in this manuscript, in this case, plane lines represent Kronecker deltas.}
    \label{fig:lncWoo}
\end{figure}
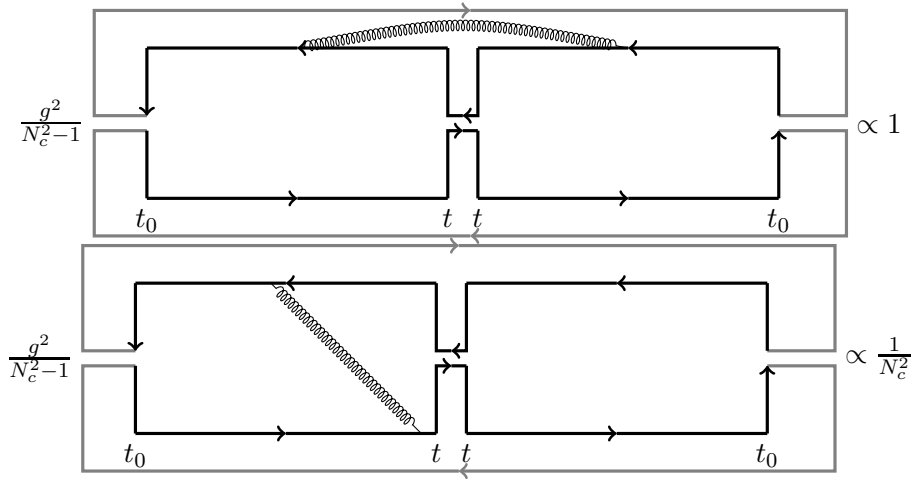

Finally, let us look at fig. \ref{fig:Wloo}. Note that any diagram connecting the upper line (the quark) with
the lower line (the antiquark) is suppressed by a factor of $\frac{1}{N_{c}^{2}}$. This is illustrated with two examples in fig. \ref{fig:lncWoo}. The physical consequence in this process is that, at leading order in this expansion,
we can consider the quark and the antiquark as uncorrelated
particles.

The reader might have noted that several simplifications based on the multiplication of Wilson lines (for example that $U^\dagger(t,t_0,\mathbf{r})U(t,t_0,\mathbf{r})=1$) were not used. This is because, although we are studying the evolution of $\rho_x(\mathbf{R},\mathbf{r},\mathbf{R},\mathbf{r},t)$, we want our discussion to illustrate what would happen in the more general $\mathbf{r}\neq\mathbf{r}'$ case. Nevertheless, we discuss these simplifications here. It can easily be seen using the results shown in fig. \ref{fig:lncWso} that
\begin{equation}
P_o(\mathbf{R},\mathbf{r},t|\rho_s)=D_{s}({\bf R},{\bf r},{\bf R},{\bf r};t_{0})-P_s(\mathbf{R},\mathbf{r},t|\rho_s)\,,
\end{equation}  
which is just an illustration that probability is conserved. Analogously, looking at fig. \ref{fig:Wloo}, we find that

\begin{equation}
P_{o}({\bf R},{\bf r},t|\rho_{o})=D_{o}({\bf R},{\bf r},{\bf R},{\bf r};t_{0})+\mathcal{O}\left(\frac{1}{N_{c}^{2}}\right)\,.
\end{equation}
This result is a combination of the conservation of probability with the fact that the transition from octets to singlets is suppressed in the large $N_c$ limit.

In summary, in the strict $\mathbf{R}=\mathbf{R}'$ and $\mathbf{r}=\mathbf{r}'$ limit, using the Fierz identity, we get the following results. Let us define
\begin{equation}
S(\mathbf{R},\mathbf{r},t-t_0)=Tr(W_{SS}^{\dagger}({\bf R},{\bf r};t,t_{0})W_{SS}({\bf R},{\bf r};t,t_{0})\rho_{l})\,.
\end{equation} 
Then we get
\begin{align}
P_s(\mathbf{R},\mathbf{r},t|\rho_s)&=S(\mathbf{R},\mathbf{r},t-t_0)D_s(\mathbf{R},\mathbf{r},\mathbf{R},\mathbf{r};t_0)\,,\\
P_o(\mathbf{R},\mathbf{r},t|\rho_s)&=(1-S(\mathbf{R},\mathbf{r},t-t_0))D_s(\mathbf{R},\mathbf{r},\mathbf{R},\mathbf{r};t_0)\,,\\
P_s(\mathbf{R},\mathbf{r},t|\rho_o)&=\frac{1-S(\mathbf{R},\mathbf{r},t-t_0)}{N_c^2-1}D_o(\mathbf{R},\mathbf{r},\mathbf{R},\mathbf{r};t_0)\,,\\
P_o(\mathbf{R},\mathbf{r},t|\rho_o)&=\frac{N_c^2-2+S(\mathbf{R},\mathbf{r},t-t_0)}{N_c^2-1}D_o(\mathbf{R},\mathbf{r},\mathbf{R},\mathbf{r};t_0)\,.
\end{align}
These equations show that the octet is much more stable than the singlet. If we start from a singlet, at a later time we may find a  singlet with zero probability (if $S=1$). Instead, if we start with an octet, the probability to find an octet later is always greater than or equal to $\frac{N_c^2-2}{N_c^2-1}$.

We can rewrite these results as transitions from $D_x(t_0)$ to $D_x(t)$ if we assume that $\rho_l\sim\tilde{\rho}_l$ (now we do not write the coordinates explicitly but they are assumed to be the same as before).
\begin{align}
D_s(t)&=S(t-t_0)D_s(t_0)+\frac{1-S(t-t_0)}{N_c^2-1}D_o(t_0)\,,\\
D_o(t)&=(1-S(t-t_0))D_s(t_0)+\frac{N_c^2-2+S(t-t_0)}{N_c^2-1}D_o(t_0)\,.
\end{align}
Note that it is fulfilled that $D_s(t)+D_o(t)=D_s(t_0)+D_o(t_0)$. If we assume that $D_s(t_0)$ and $D_o(t_0)$ are similar in size, then
\begin{align}
D_s(t)&\sim S(t-t_0)D_s(t_0)\,,\\
D_o(t)&\sim (1-S(t-t_0))D_s(t_0)+D_o(t_0)\,.
\end{align}
However, if we assume that $D_o(t_0)\sim 1$ while $D_s(t_0)\sim\frac{1}{N_c^2}$, then
\begin{align}
D_s(t)&\sim S(t-t_0)D_s(t_0)+\frac{1-S(t-t_0)}{N_c^2-1}D_o(t_0)\,,\\
D_o(t)&\sim D_o(t_0)\,.
\end{align}
These results suggest the following physical interpretation. If $D_s$ and $D_o$ are similar in size, then the evolution of the singlet is independent of the number of octets and can be modeled with an effective Hamiltonian. Moreover, the octet to singlet transition can be ignored. This evolution naturally leads to a state in which $D_s$ is much smaller than $D_o$. When $D_s$ is $N_c^2$ times smaller than $D_o$, we reach a different regime. In this case, it is the octet that evolves independently of the singlet. However, the decay of the octet into singlet is an important contribution to $D_s(t)$. These qualitative results are not far from the toy model discussed in \cite{Blaizot:2018oev}.
\section{An alternative graphical representation and the classical limit}
\label{sec:class}
In this section, we are going to show that the Wilson loops that appear in our study have some similarities with other Wilson loops that appear in the study of the Color Glass Condensate (see \cite{Blaizot:2016qgz} for a review), the dipole and the quadrupole \cite{Dominguez:2012ad}. To see this, we are going to introduce an alternative graphical representation which follows these rules:
\begin{itemize}
\item Fields on the two branches of the Schwinger-Keldysh contour are represented in different colors. Blue for fields on the upper branch of the contour and garnet for the lower one. The classical limit is then recovered graphically by ignoring the color difference. By \textit{classical}, we mean that gauge fields have the same value in the two branches of the contour. 
\item We restrict ourselves to the case $\mathbf{R}=\mathbf{R}'=\mathbf{0}$ but, unlike in the previous sections, we allow $\mathbf{r}$ and $\mathbf{r}'$ to be different. Since the three coordinates we are taking into account are not necessarily in the same plane, points in which different colors meet represent points in which the Wilson line has a cusp.
\item Time always flows from left ($t_0$) to right ($t$). The position of the particles is given from top to bottom in the following order: a quark at $\mathbf{R}+\frac{\mathbf{r}}{2}$, a quark at $\mathbf{R}'+\frac{\mathbf{r}'}{2}$, an antiquark at $\mathbf{R}'-\frac{\mathbf{r}'}{2}$ and finally an antiquark at $\mathbf{R}-\frac{\mathbf{r}}{2}$. While this ordering may seems unintuitive, it is the one that allows a direct connection with the quadrupole discussed in \cite{Dominguez:2012ad}.
\end{itemize}

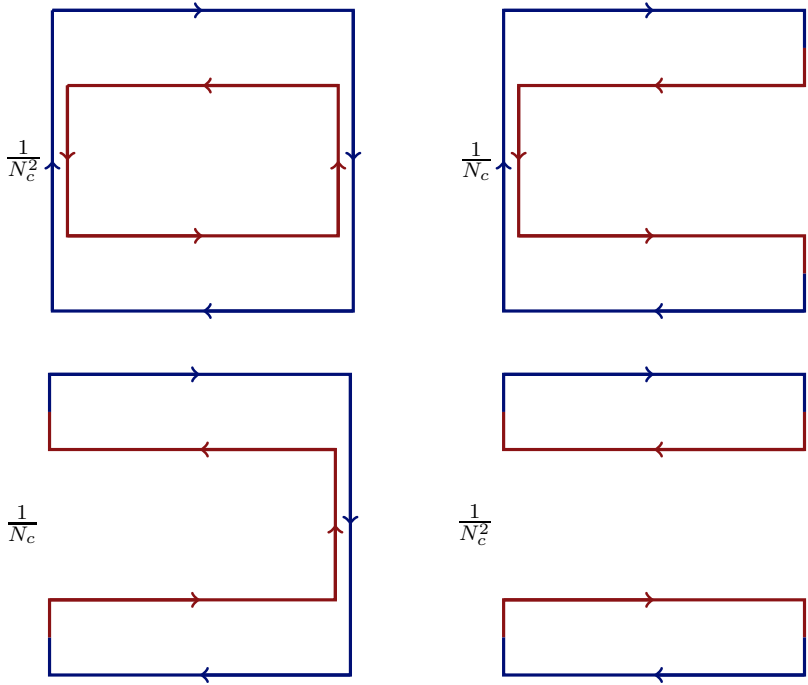
\begin{figure}
\begin{tikzpicture}
\path (0,0) node [anchor=east] {$\frac{1}{N_c^2}$};
\begin{scope}[very thick]
\draw[blau] (0,2)--(4,2)--(4,-2)--(0,-2)--(0,2);
\draw[->,blau] (0,2)--(2,2);
\draw[->,blau] (4,2)--(4,0);
\draw[->,blau] (4,-2)--(2,-2);
\draw[->,blau] (0,-2)--(0,0);
\draw[grana] (0.2,1)--(3.8,1)--(3.8,-1)--(0.2,-1)--(0.2,1);
\draw[->,grana] (0.2,1)--(0.2,0);
\draw[->,grana] (0.2,-1)--(2,-1);
\draw[->,grana] (3.8,-1)--(3.8,0);
\draw[->,grana] (3.8,1)--(2,1);
\end{scope}
\end{tikzpicture}
\hspace{1cm}
\begin{tikzpicture}
\path (0,0) node [anchor=east] {$\frac{1}{N_c}$};
\begin{scope}[very thick]
\draw[blau] (4,-1.5)--(4,-2)--(0,-2)--(0,2)--(4,2)--(4,1.5);
\draw[->,blau] (0,2)--(2,2);
\draw[->,blau] (4,-2)--(2,-2);
\draw[->,blau] (0,-2)--(0,0);
\draw[grana] (4,1.5)--(4,1)--(0.2,1)--(0.2,-1)--(4,-1)--(4,-1.5);
\draw[->,grana] (4,1)--(2,1);
\draw[->,grana] (0.2,1)--(0.2,0);
\draw[->,grana] (0.2,-1)--(2,-1);
\end{scope}
\end{tikzpicture}
\linebreak
\linebreak
\begin{tikzpicture}
\path (0,0) node [anchor=east] {$\frac{1}{N_c}$};
\begin{scope}[very thick]
\draw[blau] (0,1.5)--(0,2)--(4,2)--(4,-2)--(0,-2)--(0,-1.5);
\draw[->,blau] (0,2)--(2,2);
\draw[->,blau] (4,2)--(4,0);
\draw[->,blau] (4,-2)--(2,-2);
\draw[grana] (0,-1.5)--(0,-1)--(3.8,-1)--(3.8,1)--(0,1)--(0,1.5);
\draw[->,grana] (0,-1)--(2,-1);
\draw[->,grana] (3.8,-1)--(3.8,0);
\draw[->,grana] (3.8,1)--(2,1);
\end{scope}
\end{tikzpicture}
\hspace{1cm}
\begin{tikzpicture}
\path (0,0) node [anchor=east] {$\frac{1}{N_c^2}$};
\begin{scope}[very thick]
\draw[blau] (0,1.5)--(0,2)--(4,2)--(4,1.5);
\draw[->,blau] (0,2)--(2,2);
\draw[grana] (4,1.5)--(4,1)--(0,1)--(0,1.5);
\draw[->,grana] (4,1)--(2,1);
\draw[grana] (0,-1.5)--(0,-1)--(4,-1)--(4,-1.5);
\draw[->,grana] (0,-1)--(2,-1);
\draw[blau] (4,-1.5)--(4,-2)--(0,-2)--(0,-1.5);
\draw[->,blau] (4,-2)--(2,-2);
\end{scope}
\end{tikzpicture}
\caption{Graphical representation of the building blocks discussed in the text. The representation follows what is discussed in this section. We name the building blocks in the following way. On the top left, we find the singlet dipoles ($Sd$), on the top right, the quadrupole ($Q$), on the bottom left, the inverse quadrupole ($Iq$) and, finally, on the bottom right, the quark dipoles ($Qd$).}
\label{fig:blocks}
\end{figure}

We can identify four building blocks from which we can obtain all the transitions that we discussed previously. Two of them correspond in the classical limit to two dipoles, while the other two correspond to quadrupoles. They are represented in fig. \ref{fig:blocks}. To simplify the discussion, we have given them names which are not meant to be understood literally. In the classical limit, a quark on the lower branch of the Schwinger-Keldysh contour is analogous to an antiquark on the upper branch, and vice versa. Then, we can either form two singlets by combining each quark with an antiquark or by combining a quark on the upper branch with a quark on the lower branch (and the same for the antiquarks). Singlet dipoles would correspond to the transition of two singlets made of a quark-antiquark pair, while quark dipoles would correspond to the transition of two singlets made of a quark on the upper branch and another on the lower branch, and an antiquark on the upper branch and another on the lower branch. The quadrupole and the inverse quadrupole represent transitions between these two ways of forming singlet states. 

The conditional probabilities discussed in section \ref{sec:Conditional-probabilities-in} can be written in terms of these building blocks. A more compact way to write the results is to use the building blocks to describe the transitions from $D_x(t_0)$ to $D_x(t)$:
\begin{equation}
D_s(t)=Sd(t-t_0)D_s(t_0)+\frac{Iq(t-t_0)-Sd(t-t_0)}{N_c^2-1}D_o(t_0)\,,
\label{eq:Dsbb}
\end{equation}
\begin{align}
D_o(t)&=(Q(t-t_0)-Sd(t-t_0))D_s(t_0)\nonumber\\
&+\frac{N_c^2Qd(t-t_0)-Q(t-t_0)-Iq(t-t_0)+Sd(t-t_0)}{N_c^2-1}D_o(t_0)\,.
\label{eq:Dobb}
\end{align}
Using that the building blocks are defined such that they are of order $1$ in the large $N_c$ counting, we can directly recover some of the results of the previous section.

An interesting observation is that the singlet to octet (octet to singlet) transition is zero if the expectation value of the quadrupole (inverse quadrupole) is equal to the expectation value of the square of two singlet dipoles. From the point of view of perturbation theory, this indicates that we have to take into account diagrams in which particles traverse the cut (in other words, connect fields on different branches of the Schwinger-Keldysh contour).

One lesson that we can take from the results in this section is the following: the operators defined in section \ref{sec:Conditional-probabilities-in} can be written in the classical limit in terms of gauge invariant dipoles and quadrupoles. In this way, the issues that forced us to use two different density matrices for light quarks and gluons ($\rho_l$ and $\tilde{\rho}_l$) disappear in the classical limit.
\section{Comparison with other derivations of the evolution equation and their behavior in the large $N_c$ limit.\label{sec:Comparison-with-perturbative}}

In this section, we check if the physical consequences of our static
and large $N_{c}$ computation are compatible with state-of-the-art
perturbative derivations of the evolution equation. In particular,
we will compare with \cite{Blaizot:2017ypk,Blaizot:2018oev}
and \cite{Brambilla:2016wgg,Brambilla:2017zei} although these computations do not rely on the same assumptions. Before starting with
the comparison, let us remark that, in our derivation, the only small parameter is $\frac{1}{N_{c}}$ since we take the exact limit $M\to\infty.$ In more realistic settings,
there are other small parameters that might appear to spoil
the naive power counting based on the large $N_{c}$ limit. For example,
in the non-static case, we have to consider both bound and unbound
states. Then the results will depend on the relation between the size
of the bound states and the volume of the medium. 

To start, let us first compare with the results in \cite{Brambilla:2016wgg},
which are valid in the limit $\frac{1}{r}\gg T\gg E$ where $r$ is
the typical radius and $E$ is the binding energy. In that work, it
was found that the evolution could be written as a GKSL equation

\begin{equation}
\frac{d\rho}{dt}=-i[H,\rho]+\sum_{n}\left(C_{n}\rho C_{n}^{\dagger}-\frac{1}{2}\{C_{n}^{\dagger}C_{n},\rho\}\right)\,.
\end{equation}

In this equation, the reduced density matrix can be divided into two
pieces, one for singlet and another for octet states

\begin{equation}
\rho=\left(\begin{array}{cc}
\rho_{s} & 0\\
0 & \rho_{o}
\end{array}\right)\,.
\end{equation}

The Hamiltonian is given by 

\begin{equation}
H=\left(\begin{array}{cc}
h_{s} & 0\\
0 & h_{0}
\end{array}\right)+\frac{r^{2}\gamma}{2}\left(\begin{array}{cc}
1 & 0\\
0 & \frac{N_{c}^{2}-2}{2(N_{c}^{2}-1)}
\end{array}\right)\,,
\end{equation}

where the first term is the vacuum Hamiltonian and the parameter $\gamma$
is defined as 

\begin{equation}
\gamma=\frac{g^{2}}{6N_{c}}\Im\int_{-\infty}^{\infty}\,ds\langle\tilde{E}^{a,i}(s,{\bf 0})\tilde{E}^{a,i}(0,{\bf 0})\rangle\,.
\end{equation}
For a discussion on extractions of $\gamma$ from lattice QCD, see \cite{Brambilla:2019tpt,Eller:2019spw}. In the last equation, $\tilde{E}=\Omega E\Omega^{\dagger}$ with $E$ the chromoelectric
field and $\Omega(t,{\bf r})=U(t,-\infty;{\bf r})$. Note that regarding
the $N_{c}$ counting $\gamma$ is of order one since $g^{2}\sim\mathcal{O}\left(\frac{1}{N_{c}}\right)$.
There are two sets of collapse operators

\begin{equation}
C_{i}^{0}=\sqrt{\frac{\kappa}{N_{c}^{2}-1}}r^{i}\left(\begin{array}{cc}
0 & 1\\
\sqrt{N_{c}^{2}-1} & 0
\end{array}\right)\,,
\end{equation}

and

\begin{equation}
C_{i}^{1}=\sqrt{\frac{(N_{c}^{2}-4)\kappa}{2(N_{c}^{2}-1)}}r^{i}\left(\begin{array}{cc}
0 & 0\\
0 & 1
\end{array}\right)\,,
\end{equation}

where

\begin{equation}
\kappa=\frac{g^{2}}{6N_{c}}\Re\int_{-\infty}^{\infty}\,ds\langle\tilde{E}^{a,i}(s,{\bf 0})\tilde{E}^{a,i}(0,{\bf 0})\rangle\,.
\end{equation}
$\kappa$ is the heavy quark diffusion coefficient \cite{CasalderreySolana:2006rq,CaronHuot:2007gq}. Recent lattice QCD determinations can be found in \cite{Francis:2015daa,Brambilla:2020siz,Altenkort:2020fgs}. Note that in this case also $\kappa$ is of order one in the large
$N_{c}$ counting. 

Now that we have reproduced the evolution equations of \cite{Brambilla:2016wgg}
we are ready to compare it with the static limit results that we obtained
in section \ref{sec:Results-in-the}.
\begin{itemize}
\item The survival probability of the singlet is not suppressed by the large
$N_{c}$ limit and can be encoded in an effective non-Hermitian Hamiltonian
$h_{s}+\frac{r^{2}\gamma}{2}-i\frac{r^{2}\kappa}{2}$. This is compatible
with eq. (\ref{eq:effh}).
\item The singlet to octet transition is also of order one in the large
$N_{c}$ counting, as expected from our previous results. However,
the octet to singlet transition scales as $\frac{1}{N_{c}^{2}}$.
This can be easily seen by computing the only term in the GKSL equation
which mixes singlets and octets 
\begin{equation}
\sum_{i}C_{i}^{0}\rho C_{i}^{0\dagger}=\kappa r^{i}\left(\begin{array}{cc}
\frac{\rho_{o}}{N_{c}^{2}-1} & 0\\
0 & \rho_{s}
\end{array}\right)r^{i}\,.
\end{equation}
\item Finally, let us discuss the octet survival probability. Like that of
the singlet, it is not suppressed by large $N_{c}$ effects. This agrees with our results in section IV. Regarding the observation that the octet evolves like a pair of uncorrelated particles (see fig. \ref{fig:lncWoo} and its discussion in the text of section \ref{sec:largenc}), it is challenging to check whether this is fulfilled or not in the pNRQCD results since, in this EFT, the degrees of freedom are directly singlets and octets.
\end{itemize}
We would have arrived at the same conclusions if we compared with the more
general case presented in \cite{Brambilla:2017zei}.

Now, let us compare with the results of \cite{Blaizot:2017ypk}.
We summarize here the main assumptions of that work.
\begin{itemize}
\item The interaction Hamiltonian that describes the coupling between the heavy quarks and the medium is $H_I=-g\int\,d^3r A_0^A(\mathbf{r})n^A(\mathbf{r})$, where $n^A$ is the color current associated with the heavy particles.
\item The evolution of the density matrix is computed in the one gluon exchange approximation.
\item The response of the plasma to the perturbation caused by the heavy quarks is fast compared to the characteristic time scales of the heavy quark motion.
\end{itemize}
Under these assumptions, the following evolution is found
\begin{align}
\frac{d\rho_{s}}{dt}= & \mathcal{L}^{ss}\rho_{s}+\frac{\mathcal{L}^{so}}{N_{c}^{2}-1}\rho_{o}\,,\nonumber \\
\frac{d\rho_{o}}{dt}= & (N_{c}^{2}-1)\mathcal{L}^{os}\rho_{s}+\mathcal{L}^{oo}\rho_{o}\,.
\end{align}
In these equations,
\begin{equation}
    \mathcal{L}^{ss}=iC_F(V_{12}-V_{1'2'})+C_F(2W(0)-W_c)+\frac{C_F}{4MT}(2\nabla^2W(0)-\nabla^2W_c-\mathbf{\nabla}W_c\mathbf{\nabla}_c)\,, \label{eq:jpss}
\end{equation}
\begin{equation}
    \mathcal{L}^{so}=-C_FW^--\frac{C_F}{4MT}(\nabla^2W^-+\mathbf{\nabla}W^-\mathbf{\nabla})\,, \label{eq:jpso}
\end{equation}
\begin{equation}
    \mathcal{L}^{os}=-\frac{1}{2N_c}W^--\frac{1}{8MTN_c}(\nabla^2W^-+\mathbf{\nabla}W^-\mathbf{\nabla})\,, \label{eq:jpos}
\end{equation}
and
\begin{equation}
\begin{split}
    \mathcal{L}^{oo}&=-\frac{i}{2N_c}(V_{12}-V_{1'2'})+2C_FW(0)+\frac{1}{2N_c}W_c-\frac{N_c^2-2}{2N_c}W_a-\frac{1}{N_c}W_b\\
    &-\frac{1}{4MT}\left(\frac{N_c^2-2}{2N_c}W_a+\frac{1}{N_c}W_b\right)\,.
\end{split}
\label{eq:jpoo}
\end{equation}
The meaning of $V_{12}$, $V_{1'2'}$, $W(0)$, $W_c$, $W^-$, $W_a$ and $W_b$ can be found in \cite{Blaizot:2017ypk} and is reviewed in Appendix \ref{sec:pjp}. These functions scale like $\frac{1}{N_c}$ in the large $N_c$ counting because they are proportional to $g^2$. The following remarks are important:
\begin{itemize}
    \item $V_{12}$, $V_{1'2'}$, $W(0)$ and $W_c$ represent the effects of gluons that do not connect different branches of the Schwinger-Keldysh contour.
    \item $W_a$ and $W(0)$ represent the effect of gluons that do not connect the quark with the antiquark.
\end{itemize}
Using these remarks, we comment on the compatibility of the results of \cite{Blaizot:2017ypk} with the static limit, large $N_{c}$ conclusions of this manuscript. 
\begin{itemize}
\item The singlet to singlet transition is not suppressed by large $N_{c}$
effects and gets no contributions from terms connecting the two branches of the Schwinger-Keldysh contour.  
\item The singlet to octet transition is not suppressed while the octet
to singlet is of order $\frac{1}{N_{c}^{2}}$.
\item The octet to octet transition is not suppressed by large $N_{c}$
effects. However, we can see that the only terms that contribute to
$\mathcal{L}^{oo}$ at leading order in $N_{c}$ are those that do
not correlate the heavy quark and the antiquark. 
\end{itemize}
With this, we have finished the comparison with \cite{Blaizot:2017ypk},
we note that we would have arrived at the same conclusions if we would
have compared with \cite{Blaizot:2018oev}.
\section{Conclusions\label{sec:Conclusions}}

In this paper, we have discussed how to define the conditional probabilities
to evolve from a static pair in an octet or singlet state at a time
$t_{0}$ into an octet or singlet state at time $t$. We also argued that the study of the color transitions in the static limit is useful for the study of real quarkonium if we are in a regime in which the correlation in time between thermal fields is shorter than the inverse of the binding energy. We leave the study of a more rigorous connection to future developments in the EFT framework.  

The results
are written in terms of Wilson loops with the right transformation
properties, and they might be useful as a starting point for future investigations
using non-perturbative techniques, as for example, lattice QCD or the AdS/CFT
correspondence. Regarding lattice QCD computations, the singlet potential (both real and imaginary part) has been already computed (see \cite{Burnier:2016mxc}, for example). This provides information about the total decay width of the singlet. However, it does not give information about the differential decay width (the probability that a given singlet decays to an octet pair with some specific relative momentum). This information is included in the expectation values represented in figs. \ref{eq:Wlss} and \ref{eq:Wlso}. We are not aware of an Euclidean expectation value whose analytic continuation corresponds to what is shown in figs. \ref{eq:Wlss} and \ref{eq:Wlso}, but we do not see an a priori reason why it should not exist. This is an issue that we believe is worth further investigation given the phenomenological importance of the differential decay width of the singlet. The situation is similar regarding the expectation values represented in figs. \ref{eq:Wlos} and \ref{eq:Wloo}. This case is more challenging due to the same subtleties that appear in the computation of the octet potential and free energy \cite{Philipsen:2013ysa,Jahn:2004qr}. However, recent results \cite{Bala:2020tdt,Bazavov:2018wmo} suggest that it might be possible to overcome these difficulties. Regarding AdS/CFT, the static singlet potential has been computed by many groups (see \cite{Noronha:2009da,Hayata:2012rw} for early examples). The authors of \cite{Herzog:2002pc} gave the prescription for equilibrium Schwinger-Keldysh contour computations in the AdS/CFT correspondence. Therefore, it should be possible to compute the transition probabilities that we have discussed in the large $N_c$ limit using these techniques.

We have also discussed the large $N_c$ limit, from which we obtain the following insights:
\begin{itemize}
\item The survival probability of a singlet can be modeled with an effective Hamiltonian since diagrams that connect different branches of the Schwinger-Keldysh contour are suppressed. 
\item The effect of octet to singlet transitions in the evolution of the octet density is suppressed. The evolution of the octet can be approximated by the uncorrelated evolution of a quark and an antiquark. This is an approximation that has been used, for example, in \cite{Yao:2018nmy,Blaizot:2018oev} and which we see is justified in the large $N_c$ limit.
\item If we consider that the densities of singlets and octets are of the same size, the octet to singlet transition can be ignored. Then, the evolution of the singlet simplifies to evolution with an effective Hamiltonian. This justifies the computation of $R_{AA}$ by solving a Schr\"{o}dinger equation with an imaginary potential (see \cite{Islam:2020gdv} for a recent example) in the large $N_c$ limit as long as the population of octets is not much bigger than that of singlets. Octets evolve as an uncorrelated pair of heavy particles that is sourced by the decay of singlets into octets. The situation changes when the population of octets is much bigger than that of singlets. Then singlet to octet transition can be ignored, but octet to singlet transition must be taken into account to compute the number of singlets accurately. 
\end{itemize}
One has to be careful when extrapolating these results from the static limit to the case of real quarkonium. In that case, one has to consider that the spectrum of the singlet contains bound states (and is discrete) while that of the octet does not (and is continuous). This could lead to additional suppressions that go like the inverse of the medium volume \cite{Blaizot:2018oev}. If medium corrections are smaller than the binding energy, these effects further suppress octet to singlet transitions. However, this does not need to be so in other cases. 

We have also studied what happens when we assume that gauge fields behave classically. Then we can write the transitions in terms of building blocks which are analogous to the dipoles and quadrupoles discussed in \cite{Dominguez:2012ad} with the difference that, in our case, the Wilson lines are along time-like and space-like directions while in their case, they have light-like and space-like directions. The classical approximation could be useful to construct fully gauge invariant operators that capture part of the physics of the case in which we have an octet at the initial time. 

In this study, we have assumed a plasma formed by light quarks and gluons and only a pair of heavy particles. It would be interesting to study what happens when the number of heavy particles is increased. Notably, this is important to confirm if the large $N_c$ limit justifies the molecular chaos approximation, as our result suggests. Another possible future direction is to rigorously prove the connection between the static limit results and real quarkonium using non-relativistic EFTs.

\appendix
\section{Birdtrack notation}
\label{sec:birdtrack}
In this appendix, we introduce a diagrammatic representation suitable for the computations performed in this manuscript. It is essentially the \textit{birdtrack} notation of \cite{Cvitanovic:2008zz} (see \cite{Hidaka:2009hs} for a previous use in the field of heavy-ion physics) with the only difference that our lines represent Wilson lines in the fundamental representation instead of Kronecker deltas.
\begin{itemize}
\item Quark fields have a color subscript while antiquark fields have a color superscript. In this notation, a line represents a Wilson line with an arrow going from its superscript to its subscript. 
\item The upper part of the diagram represents the position $\mathbf{R}+\frac{\mathbf{r}}{2}$ while the lower part represents $\mathbf{R}-\frac{\mathbf{r}}{2}$. Whenever a diagram involves fields acting at different times, this will be indicated below the diagram. Therefore, vertical Wilson lines connect points at the same time and different positions while horizontal lines do the opposite.
\item A Wilson line connecting fields with the same time and coordinate are equivalent to a Kronecker delta.
\item A Kronecker delta relating indices in the adjoint representation is represented with dashed lines.   
\end{itemize}
\begin{figure}
\beginpgfgraphicnamed{fig13}
\begin{tikzpicture}
\path (0,0) node {$T^A=$}
(4,0) node {$S=\frac{1}{\sqrt{N_c}}$}
(7,0) node {$O^A=\sqrt{2}$};
\path (5.5,1) node [shape=circle,draw] {$\psi$}
(5.5,-1) node [shape=circle,draw] {$\xi^\dagger$};
\path (8.5,1) node [shape=circle,draw] {$\psi$}
(8.5,-1) node [shape=circle,draw] {$\xi^\dagger$};
\begin{scope}[very thick]
\draw[->] (1,1) -- (1,0);
\draw (1,0) -- (1,-1);
\draw[->] (5.5,0.6) -- (5.5,0);
\draw (5.5,0) --(5.5,-0.6);
\draw[->] (8.5,0.6) -- (8.5,0);
\draw (8.5,0) --(8.5,-0.6);
\end{scope}
\draw [dashed] (1,0) -- (2,0);
\draw [dashed] (8.5,0) -- (9.5,0);
\end{tikzpicture}
\endpgfgraphicnamed
\caption{Representation of the group generator $T^A$ and the fields $S$ and $O^A$ in our graphical notation.}
\label{fig:bt}
\end{figure}
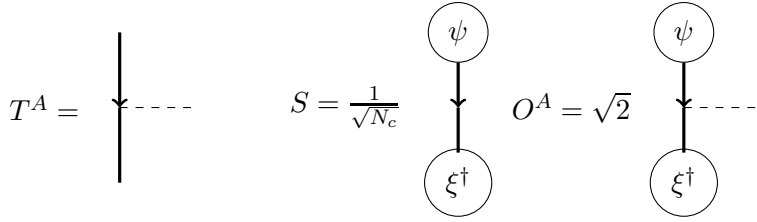
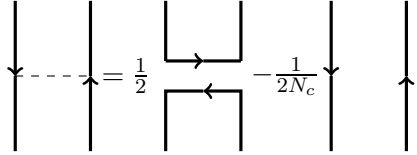
\begin{figure}
\beginpgfgraphicnamed{fig14}
\begin{tikzpicture}
\path (3,0) node[anchor=west] {$-\frac{1}{2N_c}$};
\begin{scope}[very thick]
\draw[->] (0,1)--(0,0);
\draw (0,0)--(0,-1);
\draw[->] (1,-1)--(1,0) node[anchor=west] {$=\frac{1}{2}$};
\draw (1,0)--(1,1);
\draw (2,1)--(2,0.2);
\draw[->] (2,0.2)--(2.5,0.2);
\draw (2.5,0.2)--(3,0.2);
\draw (3,0.2)--(3,1);
\draw[->] (3,-1)--(3,-0.2)--(2.5,-0.2);
\draw (2.5,-0.2)--(2,-0.2)--(2,-1);
\draw[->] (4.2,1)--(4.2,0);
\draw (4.2,0)--(4.2,-1);
\draw[->] (5.2,-1)--(5.2,0);
\draw (5.2,0)--(5.2,1);
\end{scope}
\draw[dashed] (0,0)--(1,0);
\end{tikzpicture}
\endpgfgraphicnamed
\caption{Fierz identity in our graphical notation.}
\label{fig:btF}
\end{figure}
In fig. \ref{fig:bt}, we show the group generator $T^A$ and the definition of the singlet and octet field in this notation. In fig. \ref{fig:btF}, the Fierz identity is represented in this graphical notation. Note that, for example, using the definitions of the singlet and octet field from fig. \ref{fig:bt} and the Fierz identity, it is straightforward to derive eq. (\ref{eq:on}). 
\section{Building blocks for $\mathbf{r}\neq\mathbf{r'}$.}
\label{sec:expblocks}
Here we write explicitly the building blocks that were introduced in section \ref{sec:class}. With them, it is easy to obtain the probability for any transition using eqs. (\ref{eq:Dsbb}) and (\ref{eq:Dobb}). To simplify the notation, we assume that $\mathbf{R}=\mathbf{R'}=\mathbf{0}$. We need to introduce some notation. If the density matrix of light quarks and gluons is $\rho_l$ and we denote by $|n\rangle$ an eigenvector of this matrix such that $\rho_l=\sum_n p_n|n\rangle\langle n|$, then we define $\mathit{Tr}_l$ such that
\begin{equation}
\mathit{Tr}_l(A)=\sum_n p_n\langle n|A|n\rangle\,.
\end{equation}
Similarly, $\mathit{Tr}_c$ is a color trace. We write simply $\mathit{Tr}$ when it is a combination of the two traces. To distinguish fields from different branches of the Schwinger-Keldysh contour, we write a subindex $1$ or $2$ in the corresponding Wilson lines. Additionally, we also use the color code of fig. \ref{fig:blocks}.

After fixing the notation, we give the results using the arbitrary names we introduced in section \ref{sec:class}. The singlet dipoles give
\begin{equation}
Sd(\mathbf{r},\mathbf{r}',t,t_0)=\mathit{Tr}_l(\mathcal{P}{\color{grana} W_{SS,2}^\dagger(\mathbf{0},\mathbf{r}';t,t_0)}{\color{blau} W_{SS,1}(\mathbf{0},\mathbf{r};t,t_0)})\,.
\end{equation}
The quadrupole gives
\begin{align}
Q(\mathbf{r},\mathbf{r}',t,t_0)&=\frac{1}{N_c}\mathit{Tr}\left(\mathcal{P}{\color{grana}\phi_2(\mathbf{0},-\mathbf{r}'/2;t)U_2(t,t_0;-\mathbf{r}'/2)\phi_2(-\mathbf{r}'/2,\mathbf{r}'/2;t_0)U_2^\dagger(t,t_0;\mathbf{r}'/2)}\times\right.\nonumber\\
&\times\left.{\color{grana}\phi_2(\mathbf{r}'/2,\mathbf{0};t)}{\color{blau}\phi_1(\mathbf{0},\mathbf{r}/2;t)U_1(t,t_0,\mathbf{r}/2)\phi_1(\mathbf{r}/2,-\mathbf{r}/2;t_0)U_1^\dagger(t,t_0;-\mathbf{r}/2)}\times\right.\nonumber\\
&\left.\times{\color{blau}\phi_1(-\mathbf{r}/2,\mathbf{0};t)}\right)\,.
\end{align}
The inverse quadrupole gives
\begin{align}
Iq(\mathbf{r},\mathbf{r}',t,t_0)&=\frac{1}{N_c}\mathit{Tr}\left(\mathcal{P}{\color{grana}\phi_2(\mathbf{0},\mathbf{r}'/2;t_0)U_2^\dagger(t,t_0;\mathbf{r}'/2)\phi_2(\mathbf{r}'/2,-\mathbf{r}'/2;t)U_2(t,t_0;-\mathbf{r}'/2)}\times\right.\nonumber\\
&\times\left.{\color{grana} \phi_2(-\mathbf{r}'/2,\mathbf{0};t_0)}{\color{blau} \phi_1(\mathbf{0},-\mathbf{r}/2;t_0)U_1^\dagger(t,t_0;-\mathbf{r}/2)\phi_1(-\mathbf{r}/2,\mathbf{r}/2;t)U_1(t,t_0;\mathbf{r}/2)}\times\right.\nonumber\\
&\times\left.{\color{blau} \phi_1(\mathbf{r}/2,\mathbf{0};t_0)}\right)\,.
\end{align}
Finally, the quark dipoles give
\begin{align}
Qd(\mathbf{r},\mathbf{r}',t,t_0)&=\frac{1}{N_c^2}\mathit{Tr}_l\left(\mathcal{P}\mathit{Tr}_c\left({\color{grana} \phi_2(\mathbf{0},-\mathbf{r}'/2;t)U_2(t,t_0;-\mathbf{r}'/2)\phi_2(-\mathbf{r}'/2,\mathbf{0};t_0)}{\color{blau} \phi_1(\mathbf{0},-\mathbf{r}/2;t_0)}\times\right.\right.\nonumber\\
&\times\left.\left.{\color{blau} U_1^\dagger(t,t_0,-\mathbf{r}/2)\phi_1(-\mathbf{r}/2,\mathbf{0};t)}\right)\mathit{Tr}_c\left({\color{grana} \phi_2(\mathbf{0},\mathbf{r}'/2;t_0)U_2^\dagger(t,t_0;\mathbf{r}'/2)\phi_2(\mathbf{r}'/2,\mathbf{0};t)}\times\right.\right.\nonumber\\
&\times\left.\left.{\color{blau} \phi_1(\mathbf{0},\mathbf{r}/2;t)U_1(t,t_0;\mathbf{r}/2)\phi_1(\mathbf{r}/2,\mathbf{0};t_0)}\right)\right)\,.
\end{align}
\section{Review of the definitions introduced in \cite{Blaizot:2017ypk}}
\label{sec:pjp}
In this appendix we define the quantities used in eqs. (\ref{eq:jpss}) to (\ref{eq:jpoo}). They are all related with correlators of the temporal gauge field $A_0$. First, let us define $\Delta$ such that
\begin{equation}\label{correlators}
\begin{split}
\langle {\rm T}[A_0^a(t_1,\mathbf{x}) A_0^b(t_1', \mathbf{x}')] \rangle_0=-i\delta^{ab}\Delta(t_1-t_1',\mathbf{x}-\mathbf{x}')\nonumber\\
 \langle\tilde {\rm T}[A_0^a(t_2,\mathbf{x}) A_0^b(t_2' ,\mathbf{x}')] \rangle_0=-i\delta^{ab}\tilde\Delta(t_2-t_2',\mathbf{x}-\mathbf{x}')\nonumber\\
 \langle {\rm T}_C  A_0^a(t_2,\mathbf{x}')A_0^b(t_1,\mathbf{x})  \rangle_0=\delta^{ab}\Delta^>(t_2-t_1,\mathbf{x}'-\mathbf{x})\nonumber\\
 =\delta^{ab}\Delta^<(t_1-t_2,\mathbf{x}-\mathbf{x}').
 \end{split}
 \end{equation}
It is convenient to switch to a mixed representation in which $\Delta$ depends on frequency and position. Then we define $V$ and $W$, respectively, as
\begin{equation}
V(\mathbf{r})=-\Delta^R(\omega=0,\mathbf{r}), \qquad W(\mathbf{r})=-\Delta^<(\omega=0,\mathbf{r}).
\end{equation}
Physically, $V$ represents the part of the interaction with the $A_0$ field is part of the unitary evolution of quarkonium's wave function. On the other hand, $W$ contributes to dissipation and is related with the imaginary part of the potential.

In \cite{Blaizot:2017ypk}, the position of the heavy quark (antiquark) on the left of the density matrix is denoted by $\mathbf{r}_1$ ($\mathbf{r}_2$) and the position on the right of the density matrix by $\mathbf{r}_1'$ ($\mathbf{r}_2'$). Then, for example
\begin{equation}
V_{12}=V(\mathbf{r}_1-\mathbf{r}_2)\,.
\end{equation}
Similarly for $V_{1'2'}$ and all other possible combinations. The same notation is used with $W$. Using this, we can define
\begin{equation}
\begin{split}
W_a\equiv W_{11'}+W_{22'}, \qquad W_b\equiv W_{21'}+W_{12'}\,,\\
W_c\equiv W_{12}+W_{1'2'},\qquad W^\pm \equiv W_a\pm W_b\,.
\end{split}
\end{equation}
\begin{acknowledgments}
We thank Jean-Paul Blaizot, Nora Brambilla, Joan Soto, Antonio Vairo and Peter Vander Griend for their comments on earlier versions of this manuscript. This work was supported by \textit{Ministerio de Ciencia e Innovacion} of Spain under project FPA2017-83814-P and Maria de Maetzu Unit of Excellence MDM-2016- 0692, by Xunta de Galicia and FEDER. 
\end{acknowledgments}

\bibliography{staticnc}

\end{document}